\def\DIRvalue{Zarembo}
\def\IDvalue{ZA}
\def\titlevalue{Localization and AdS/CFT Correspondence}
\def\authorvalue{Konstantin Zarembo}
\def\shortauthorvalue{\authorvalue}
\def\addressvalue{Nordita, KTH Royal Institute of Technology and Stockholm University,
Roslagstullsbacken 23, SE-106 91 Stockholm, Sweden\\
Department of Physics and Astronomy, Uppsala University, SE-751 08 Uppsala, Sweden\\
\tt zarembo@nordita.org}

\def\abstractvalue{ An interplay between localization and holography is
  reviewed with the emphasis on the $AdS_5/CFT_4$ correspondence.}

\def\preprintvalue{\texttt{NORDITA-2016-30} \\
\texttt{UUITP-08/16}}

\ifx\ifLONG\undefined
\documentclass[a4paper,11pt]{article}

\newcommand{\chapterauthor}[1]{
\begin{center}
{\bf \normalsize  #1}
\end{center}
}

\newcommand{\chapteraddress}[1]{
\begin{center}
{ \small \it \addressvalue}
\end{center}
}

\newcommand{\chapterabstract}[1]{
\vspace{\baselineskip}
\begin{center}
\textbf{\small Abstract}
\end{center}
#1}

\newcommand{\chapterheader}{

\chapter[\titlevalue{}  (by \shortauthorvalue)]{\titlevalue}
\label{Chapter\IDvalue}
\chapterauthor{\authorvalue}
\chapteraddress{\addressvalue}
\chapterabstract{\abstractvalue}
\tightmtctrue
\minitoc
}

\newcommand{\documentheader}{
\begin{flushright} \small
  \preprintvalue
 \end{flushright}

\begin{center}
{\bf \Large \titlevalue}
\end{center}

\chapterauthor{\authorvalue}
\chapteraddress{\addressvalue}
\chapterabstract{\abstractvalue}

\medskip

This is a contribution to the review volume ``Localization techniques
in quantum field theories'' (eds. V.~Pestun and M.~Zabzine) which
contains 17 Chapters available at \cite{ContributionSummary}

\tableofcontents
}

\newcommand{\ifvolume}[2]{\ifx\ifLONG\undefined#2\else#1\fi}

\newcommand{\documentfinish}{
\ifx\ifLONG\undefined
\bibliographystyle{bibreview} 
\bibliography{\IDvalue,review}  
\end{document}
\else
\addcontentsline{toc}{section}{References}
\input{\DIRvalue/\IDvalue.bbl}
\fi
}

\newcommand{\documentfinishBBL}{
\addcontentsline{toc}{section}{References}
\ifx\ifLONG\undefined

\end{document}
\else

\fi
}

\ifx\ifLONG\undefined
\def\volcite#1{Contribution \cite{Contribution#1}}
\else 
\def\volcite#1{Chapter \ref{Chapter#1}}
\fi

\newcommand{\ContributionSummaryBibItemReference}
{
\bibitem{ContributionSummary}
V.~Pestun and M.~Zabzine, eds., {\em Localization techniques in quantum field
  theory}, vol.~xx.
\newblock Journal of Physics A, 2016.
\newblock \href{http://arxiv.org/abs/1608.02952}{{\tt 1608.02952}}.
\newblock \url{https://arxiv.org/src/1608.02952/anc/LocQFT.pdf},
  \url{http://pestun.ihes.fr/pages/LocalizationReview/LocQFT.pdf}.
}

 \usepackage{amssymb,graphicx}
 \usepackage{amsmath}

 \usepackage[small]{subfigure}

 \usepackage{hyperref}

 \usepackage{ytableau}

\usepackage{simplewick}
 \usepackage{fullpage}

\numberwithin{equation}{section}

\begin{document}
\thispagestyle{empty}

\documentheader

\else

\chapterheader

\fi

\section{Introduction}\label{ZAsec:Introduction}

The holographic duality  \cite{ZAMaldacena:1998re,ZAGubser:1998bc,ZAWitten:1998qj} can be understood as a precise string
reformulation of the large-$N$ expansion \cite{ZAHooft:1973jz}. Which gauge theories (perhaps all?) have exact holographic duals is an interesting open problem.  
 So far the classical gravity approximation has been the most useful holographic setup. This approximation is restricted to the regime of a very large 't~Hooft coupling and is obviously difficult to access by ordinary methods of quantum field theory. 
 Any exact result that can be reliably extended to strong coupling and consequently confronted with the predictions of holography is very valuable in this respect.
A number of such results can be obtained with the help of supersymmetric localization  \cite{ZAPestun:2007rz}.  
Localization  is thus instrumental in connecting holography with down-to-earth quantum field theory calculations, and opens a window onto genuine non-perturbative dynamics of gauge fields.

The aim of these notes is to review the large-$N$ expansion of localization formulas, with the aim to connect them to string theory and holographic duality.
The review almost exclusively deals with the maximally supersymmetric $\mathcal{N}=4$ super-Yang-Mills (SYM) theory in four dimensions, apart from a rather cryptic discussion of its less supersymmetric non-conformal cousins. According to the AdS/CFT correspondence,  $\mathcal{N}=4$ SYM is equivalent to string theory on $AdS_5\times S^5$ \cite{ZAMaldacena:1998re}. The AdS/CFT correspondence is the first and best studied model of holographic duality. 
While holographic uses  of localization  are  not restricted to this setup, other cases are extensively covered elsewhere. Localization in three dimensions and its applications to  the $AdS_4/CFT_3$ duality are covered in \cite{ZAMarino:2011nm} and in \volcite{MA}. Massive theories in four dimensions are treated in more detail in  \cite{ZARusso:2013sba}. An overview of early developments in $\mathcal{N}=4$ SYM, with applications to the $AdS_5/CFT_4$ correspondence, can be found in \cite{ZASemenoff:2002kk}.

\section{$\mathcal{N}=4$ Super-Yang-Mills theory}\label{ZAsec:N=4SYM}

The field content of the $\mathcal{N}=4$ SYM  \cite{ZAGliozzi:1976qd,ZABrink:1976bc} consists of the gauge potentials $A_\mu $, six scalars $\Phi _I$ and four Majorana fermions $\Psi _{\alpha A}$, all in the adjoint representation of the gauge group. The scalars and fermions transform in the $\mathbf{6}$ and $\mathbf{4}$ (vector and spinor) representations of the $SO(6)$ R-symmetry group. The Lagrangian of $\mathcal{N}=4$ SYM can be obtained by dimensionally reducing $D=10$, $\mathcal{N}=1$  Yang-Mills theory to four dimensions  \cite{ZAGliozzi:1976qd,ZABrink:1976bc}. The ten-dimensional origin of the theory is reflected in its field content: the bosons $(A_\mu ,\Phi _I)$ combine into the ten-dimensional vector potential and $\Psi _{\alpha A}$ are components of a single ten-dimensional Majorana-Weyl spinor. The action takes a rather concise form in the 10d notations, which is very useful for practical purposes:
\begin{eqnarray}\label{ZAaction}
 S&=&\frac{1}{g_{\rm YM}}\int_{}^{}d^4x\,\mathop{\mathrm{tr}}
 \left[
 -\frac{1}{2}\,F_{\mu \nu }^2+\left(D_\mu \Phi _I\right)^2+\frac{1}{2}\,\left[\Phi _I,\Phi _J\right]^2
 \right.
\nonumber \\
&&\left.\vphantom{\frac{1}{2}\,F_{\mu \nu }^2}
+i\bar{\Psi }\gamma ^\mu D_\mu \Psi 
 +\bar{\Psi }\gamma ^I\left[\Phi _I,\Psi \right]
 \right],
\end{eqnarray}
The Dirac matrices $\gamma ^M=(\gamma ^\mu ,\gamma ^I)$ form the  ten-dimensional Clifford algebra,  the fermions satisfy $\gamma ^{11}\Psi =\Psi $ and $\bar{\Psi }=\Psi ^tC$, where $\Gamma ^{11}$ and $C$ are the ten-dimensional chirality and charge-conjugation matrices, respectively.  One can choose $\gamma ^I=\gamma ^5\Gamma ^I$, where $\Gamma ^I$ are the six-dimensional Dirac matrices, and assume that $\gamma ^\mu $ only act on the 4d spinor indices $\alpha $ and $\Gamma ^I$ only act the R-symmetry indices $A$.

The $\mathcal{N}=4$ supersymmetry is the largest possible in non-gravitational theories in four dimensions and is powerful enough to protect the coupling constant  $g_{\rm YM}$ from renormalization. The coupling therefore does not run with the energy scale and as a consequence the classical conformal invariance of the SYM action is not broken by quantum corrections.  The $\mathcal{N}=4$ SYM therefore constitutes a continuous family of conformal field theories parameterized by the gauge coupling $g_{\rm YM}$, the theta angle (which we have so far set to zero), and the gauge group, here taken to be $U(N)$.

The AdS/CFT duality relates $\mathcal{N}=4$ SYM to type IIB superstring theory on $AdS_5\times S^5$   \cite{ZAMaldacena:1998re,ZAGubser:1998bc,ZAWitten:1998qj}. The duality is naturally formulated within the large-$N$ expansion and is especially simple in the large-$N$ limit, in which the 't~Hooft coupling
\begin{equation}
 \lambda =g_{\rm YM}^2N
\end{equation}
is kept fixed while $N\rightarrow \infty $. The string coupling and the dimensionless string tension are related to the parameters of the SYM theory as \cite{ZAMaldacena:1998re}
\begin{equation}\label{ZAindenparAdS/CFT}
 g_s=\frac{\lambda }{4\pi N}\,\qquad T\equiv \frac{R_{\rm AdS}^2}{2\pi \alpha '}=\frac{\sqrt{\lambda }}{2\pi }\,.
\end{equation}
The planar (infinite-$N$) limit of the gauge theory thus maps  to the non-interacting string theory, which is still a fairly complicated quantum-me\-cha\-ni\-cal system.
The string tension $T$ is defined as the coupling multiplying the string action. The radius of AdS $R_{\rm AdS}$ and the string length $\sqrt{\alpha '}$ can only appear in this combination, and never alone, because any dimensionful parameter is forbidden by scale invariance.

The AdS metric, written in the units where the AdS radius $R_{\rm AdS}$ is set to one, is given by
\begin{equation}\label{ZAPoincare}
 ds^2=\frac{dx_\mu ^2+dz^2}{z^2}\,.
\end{equation}
The holographic radial coordinate $z$ ranges from zero to infinity. The gauge-theory observables are located at the boundary of AdS at $z=0$. There is a precise map between correlation functions in the SYM theory and string amplitudes  in $AdS_5\times S^5$ (with sources at the boundary). Moreover, when $\lambda \gg 1$ the radius of AdS is large in the string units and the string amplitudes can be approximated by gravitational perturbations classically propagating on the AdS background \cite{ZAWitten:1998qj}. The holographic duality is oftentimes identified with this simplified setup.

\section{Circular Wilson loop}\label{ZAsec:CircularWL}

One of the operators with a well established holographic dual and which at the same time can be computed by localization, is the Wilson loop, defined as \cite{ZAMaldacena:1998im}
\begin{equation}\label{ZAWLoop}
 W_{\mathcal{R}}(C,\mathbf{n})=\left\langle \mathop{\mathrm{tr}}\nolimits_\mathcal{R}{\rm P}\exp\left[
 \oint_Cds\left(i\dot{x}^\mu A_\mu +|\dot{x}|n^I\Phi _I\right)
 \right]\right\rangle.
\end{equation}
The Wilson loop is characterized by a contour $C=\{x^\mu (s)|s\in(0,2\pi )\}$ in the four-dimensional space-time (we concentrate on space-like Wilson loops, for which  $\dot{x}^2>0 $ in the $-+++$ metric), a contour on $S^5$ (parameterized by a six-dimensional unit vector $n^I$), and representation $\mathcal{R}$ of the $U(N)$ gauge group. For the defining representation ($\mathcal{R}=\Box$), the representation label will be omitted.

An important property of the Wilson loop operator is its local invariance under supersymmetry transformations. The general supersymmetry variation of the Wilson loop is
\[
 \delta _\epsilon W=\left\langle 
 \mathop{\mathrm{tr}}\nolimits_\mathcal{R}{\rm P}
 \oint d\tau \,\bar{\epsilon }\left(i\dot{x}^\mu \gamma _\mu +|\dot{x}|\gamma ^5n^I\Gamma _I\right)\Psi 
 \exp\left[
 \int_\tau ^{\tau+2\pi } ds\left(i\dot{x}^\mu A_\mu +|\dot{x}|n^I\Phi _I\right)
 \right]
 \right\rangle.
\]
As soon as the 6-vector $n^I$ has the unit norm the combination of the Dirac matrices that enters the  variation is degenerate, because it squares to zero:
\[
 \left(i\dot{x}^\mu \gamma _\mu +|\dot{x}|\gamma ^5n^I\Gamma _I\right)^2=0.
\]
Alternatively, the spinor product in the variation can be written as
\begin{equation}
 \bar{\epsilon }\left(i\dot{x}^\mu \gamma _\mu +|\dot{x}|\gamma ^5n^I\Gamma _I\right)\Psi 
 =i\bar{\epsilon }\mathcal{P}^+\dot{x}^\mu \gamma _\mu \Psi ,
\end{equation}
where
\begin{equation}\label{ZAbasicprojector}
 \mathcal{P}^\pm=1\pm i\,\frac{\dot{x}^\mu  }{|\dot{x}|}\,\gamma _\mu
 \gamma ^5n^I\Gamma _I,
\end{equation}
are orthogonal half-rank projectors. 

Choosing
\begin{equation}\label{ZAsuperspinor}
 \bar{\epsilon }=\bar{\epsilon }_0\mathcal{P}^-,
\end{equation}
forces the supervariation of the Wilson loop to vanish. The projectors $\mathcal{P}^\pm$ however depend on the position on the contour, through the velocity vector $\dot{x}^\mu $, and so does the  parameter of variation $\bar{\epsilon }$.  For the Wilson loop to be a real supersymmetric invariant, $\bar{\epsilon }$ must be constant. An example  is the straight line, for which the projectors $\mathcal{P}^\pm$ are constant. As a consequence the straight Wilson line preserves half of the supersymmetry and does not receive any quantum corrections due to supersymmetry protection. A more general construction  allows for arbitrary  space-time dependence, but involves a non-trivial contour on $S^5$  correlated with the space-time contour $C$ \cite{ZAZarembo:2002an}. 

The local super-invariance  is not an honest symmetry of the action, but it is sufficient to protect Wilson loops from divergent quantum corrections.
The UV divergences arise from small-scale fluctuations of quantum fields, and 
at short distances any smooth contour resembles the straight line, which is supersymmetric. Supersymmetry is not sufficient to cancel all quantum corrections for arbitrary Wilson loops, but it makes them UV finite. 

An interesting intermediate case between completely trivial supersymmetric Wilson loops and too complicated generic observables are Wilson loops
 invariant under superconformal transformations. They are not entirely protected from quantum corrections, but superconformal invariance entails massive cancellations and leaves behind a relatively simple result, that sometimes can be computed by localization of the path integral. 

The supersymmetry variation is a superconformal transformation if the spinor $\bar{\epsilon }$ is a linear function of $x^\mu $:
\begin{equation}\label{ZAlinearspinor}
 \bar{\epsilon }=\bar{\eta }+ \bar{\chi }\,x^\mu\gamma _\mu  ,
\end{equation}
where $\bar{\eta }$ and $\bar{\chi }$ are arbitrary, coordinate-independent 10d Majorana-Weyl spinors. Superconformally invariant Wilson loops can be completely classified \cite{ZADymarsky:2009si}. We begin with the simplest case -- the circular loop. The expectation value for the circle can be computed exactly and reduces to a zero-dimensional Gaussian matrix model, as was initially conjectured on account of  perturbative calculations \cite{ZAErickson:2000af,ZADrukker:2000rr} and then proved by computing the path integral of $\mathcal{N}=4$ SYM  with the help of localization \cite{ZAPestun:2007rz}.

Let us show that the circular Wilson loop is invariant under half of superconformal transformations.
For the circle in the $(34)$ plane,
\begin{equation}\label{ZAgamma-id}
 \dot{x}^a =\varepsilon ^{ab }x_b ,
 \qquad 
 \gamma _a \gamma ^5=\varepsilon _{ab }\gamma ^0\gamma ^1\gamma ^b ,
\end{equation}
where indices $a$ and $b$ take values $3$ and $4$. 
Taking these identities into account, the projectors (\ref{ZAbasicprojector}) can be brought to the following form:
\begin{equation}
 \mathcal{P}^\pm=1\pm i\gamma ^0\gamma ^1n^I\Gamma _Ix^a \gamma _a .
\end{equation}
On a contour in the $(34)$ plane $x^a\gamma _a=x^\mu \gamma _\mu $, and
 the spinor (\ref{ZAsuperspinor}) then has the requisite form (\ref{ZAlinearspinor}) for any constant $\bar{\epsilon }_0$. The circular Wilson loop therefore is 1/2 BPS,  commuting with 8 superconformal generators. 
 
 Another way to see that the circle preserves half of the superconformal symmetry is to notice that it can be mapped to a straight line by a conformal transformation. The expectation values of the circle and the straight line, however, are different, which can be understood as an anomaly associated with the boundary conditions at infinity \cite{ZADrukker:2000rr,ZASemenoff:2002kk}.

When the theory is compactified on $S^4$, the BRST generator used for localization of the path integral \cite{ZAPestun:2007rz} is among the supersymmetries preserved by the Wilson loop that runs along the big circle of the sphere. In a conformal theory such as $\mathcal{N}=4$ SYM correlation functions on the sphere are equivalent to those on the plane, and therefore an expectation value of the circular Wilson loop (be it on $\mathbb{R}^4$ or on $S^4$) can be computed by localization \cite{ZAPestun:2007rz}. 

The path integral on $S^4$ localizes to zero modes of one of the scalar fields (for consistency, this has to be the same scalar that appears in the Wilson loop operator), and the partition function  reduces to the Gaussian Hermitian matrix model:
\begin{equation}\label{ZAN=4MM}
 Z=\int_{}^{}d^{N^2}\Phi \,\,{\rm e}\,^{-\frac{8\pi ^2N}{\lambda }\,\mathop{\mathrm{tr}}\Phi ^2}.
\end{equation}
The matrix-model action  originates from the $\mathcal{R}\mathop{\mathrm{tr}}\Phi^2 $ coupling to the curvature of the sphere, which is necessary to maintain the  supersymmetry on  $S^4$. The big-circle Wilson loop maps onto the exponential average in this simple matrix model \cite{ZAErickson:2000af,ZADrukker:2000rr,ZAPestun:2007rz}:
\begin{equation}\label{ZAMMWL}
 W_{\mathcal{R}}(C_{\rm circle})=\left\langle \mathop{\mathrm{tr}}\nolimits_{\mathcal{R}}\,{\rm e}\,^{2\pi \Phi } \right\rangle.
\end{equation}
The details of the path-integral localization that leads to this formula can be found in \volcite{HO} or in \cite{ZAPestun:2007rz}.

\begin{figure}[t]
\begin{center}
 \centerline{\includegraphics[width=12cm]{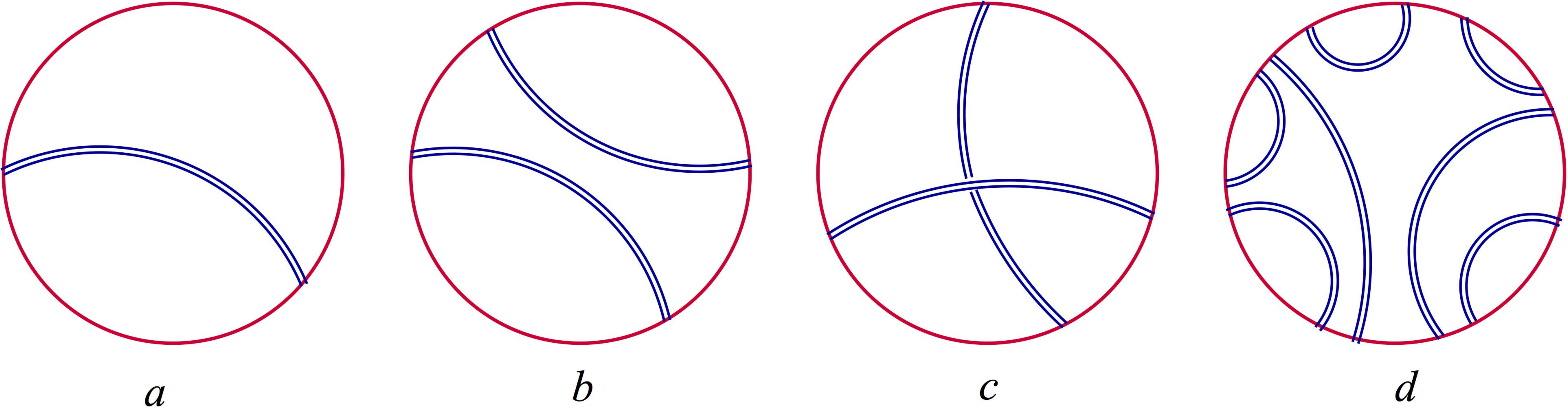}}
\caption{\label{ZAladfig}\small Rainbow diagrams that contribute to the expectation values of the circular Wilson loop. Each line is a sum of gluon and scalar propagators.}
\end{center}
\end{figure}

The same result can be derived by resumming the Feynman diagrams of perturbation theory \cite{ZAErickson:2000af,ZADrukker:2000rr}. Consider the first perturbative correction  (diagram~a in fig.~\ref{ZAladfig}):
\begin{equation}\label{ZA1-loop}
\frac{1}{N}\, W(C)=1+\,\frac{\lambda }{16\pi ^2}\oint_Cds_1\,\oint_Cds_2\,\,
 \frac{|\dot{x}_1|\,|\dot{x}_2|-\dot{x}_1\cdot \dot{x}_2}{(x_1-x_2)^2}+\ldots 
\end{equation}
The first term in the integral comes from the scalar exchange and the second from the vector propagator. For the circle in the canonical parameterization, the numerator equals $1-x_1\cdot x_2$, and the denominator $(x_1-x_2)^2=2-2x_1\cdot x_2$.
The sum of the scalar and vector exchanges combines into a constant, equal to $\lambda /16\pi ^2$. We can regard this constant as the propagator of an effective zero-dimensional field theory (\ref{ZAN=4MM}).  The two diagrams that contribute at the next order are planar and non-planar rainbow graphs, diagrams b and c in fig.~\ref{ZAladfig}. Other one-loop diagrams  appear to cancel among themselves \cite{ZAErickson:2000af}. All propagators in the rainbow graphs are effectively constant and the result is again the same as the second-order perturbation theory in the matrix model. The zero-dimensional average (\ref{ZAMMWL}) is a combinatorial tool to  generate the sum of rainbow diagrams. The diagrams with internal vertices cancel among themselves and never contribute at any order in perturbation theory \cite{ZADrukker:2000rr}. 

The expectation value (\ref{ZAMMWL}), being zero-dimensional and Gaussian, can be calculated exactly \cite{ZADrukker:2000rr}:
\begin{equation}\label{ZALaguerre}
 W(C_{\rm circle})=\,{\rm e}\,^{\frac{\lambda }{8N}}L^1_{N-1}\left(-\frac{\lambda }{4N}\right),
\end{equation}
where $L^m_n(x)$ are the Laguerre polynomials. In order to connect to the  AdS/CFT duality, we need to take  the large-$N$ limit, which amounts to summing planar rainbow graphs. A typical diagram of this type is shown in fig.~\ref{ZAladfig}d. The large-$N$ result can be extracted from (\ref{ZALaguerre}), but  it is instructive to compute the planar expectation value of the circular loop 
by  standard methods of random matrix theory \cite{ZABrezin:1977sv}, without using the exact result.

The matrix integral (\ref{ZAN=4MM}) can be reduced to eigenvalues:
\begin{equation}\label{ZAeigenvalueint}
 Z=\int_{}^{}d^N a\,\prod_{i<j}^{}\left(a_i-a_j\right)^2\,{\rm e}\,^{-
 \frac{8\pi ^2N}{\lambda }\,\sum\limits_{i}^{}a_i^2
 }.
\end{equation}
At large $N$, the  saddle-point approximation becomes exact (because the action is $\mathcal{O}(N^2)$ and there are only $\mathcal{O}(N)$ integration variables). The distribution of eigenvalues that minimizes the effective action satisfies the following set of equations:
\begin{equation}
 \frac{1}{N}\sum_{j\neq i}^{}\frac{1}{a_i-a_j}=\frac{8\pi ^2}{\lambda }\,a_i.
\end{equation}
This is an equilibrium condition for $N$ particles with logarithmic pairwise repulsion in the common harmonic potential. The  tendency of eigenvalues to fall into the bottom of the potential is counteracted by repulsion, which causes a finite spread of the eigenvalue distribution. In the thermodynamic (large-$N$) limit the distribution is characterized by a continuous density:
\begin{equation}
 \rho (x)=\frac{1}{N}\sum_{i}^{}\delta \left(x-a_i\right).
\end{equation}
In terms of the density, the saddle-point equations take a form of a singular integral equation:
\begin{equation}
 \int_{-\mu }^{\mu }\frac{dy\,\rho (y)}{x-y}=\frac{8\pi ^2}{\lambda }\,x,
 \qquad x\in(-\mu ,\mu ).
\end{equation}
 For $\mu $ fixed this equation has a unique solution, provided that $\rho (\pm \mu )=0$ \cite{ZAGakhov}. The maximal eigenvalue $\mu $ is then determined by the normalization condition. Altogether, the eigenvalue distribution takes the well-known Wigner form:
\begin{equation}\label{ZAWigner}
 \rho (x)=\frac{2}{\pi \mu ^2}\,\sqrt{\mu ^2-x^2}
\end{equation}
with 
\begin{equation}\label{ZAmuGauss}
 \mu =\frac{\sqrt{\lambda }}{2\pi }\,.
\end{equation}

The circular Wilson loop can be calculated from (\ref{ZAMMWL}):
\begin{equation}\label{ZAintMMWL}
 \frac{1}{N}\,W(C_{\rm circle})=\int_{-\mu }^{\mu }dx\,\rho (x)\,{\rm e}\,^{2\pi x},
\end{equation}
which gives \cite{ZAErickson:2000af}:
\begin{equation}\label{ZABessel}
 \frac{1}{N}\, W(C_{\rm circle})=\frac{2}{\sqrt{\lambda }}\,I_1\left(\sqrt{\lambda }\right).
\end{equation}
where $I_\nu (x)$ is the modified Bessel function. The appearance of the square root of $\lambda $  here is to some extent fictitious, because $I_1(x)$ is an odd function and the weak-coupling expansion goes in the powers of $\lambda $ as it should. 

 But at large argument the Bessel function has an essential singularity and expands in an asymptotic, non-Borel-summable 
series in $1/\sqrt{\lambda }$. To the leading order:
\begin{equation}\label{ZAStronglambdacircle}
 \frac{1}{N}\,W(C_{\rm circle})\stackrel{\lambda \rightarrow \infty}{\simeq }\sqrt{\frac{2}{\pi }}\lambda ^{-\frac{3}{4}}\,{\rm e}\,^{\sqrt{\lambda }}.
\end{equation}
Now the square root of $\lambda $ appears for real, as actually expected, because the $\hbar$ of the string sigma-model according to (\ref{ZAindenparAdS/CFT}) is $2\pi /\sqrt{\lambda }$, such that $1/\sqrt{\lambda }$ plays the r\^ole of the loop counting parameter on the worldsheet. 

We now have an explicit result at strong coupling at our disposal, computed directly from the path integral of $\mathcal{N}=4$ SYM. According to the AdS/CFT duality, the Wilson loop expectation value maps to a disc amplitude in string theory \cite{ZAMaldacena:1998im,ZARey:1998ik}:
\begin{equation}\label{ZAWinstring}
 W(C,\mathbf{n})=\int_{}^{}\mathcal{D}h_{ab}\mathcal{D}X^M\mathcal{D}\theta^\alpha  \,{\rm e}\,^{-\frac{\sqrt{\lambda }}{2\pi }\,S_{\rm str}\left[h_{ab},X^M,\theta^\alpha  \right]},
\end{equation}
where $h_{ab}$, $X^M$ and $\theta^\alpha  $ are the worldsheet metric, the string embedding coordinates  and the worldsheet fermions. The full string action is known explicitly \cite{ZAMetsaev:1998it}, but for our classical analysis the bosonic part of the sigma-model in $AdS_5$ will suffice:
\begin{equation}
 S_{\rm str}=\frac{1}{2}\int_{}^{}d\tau \,d\sigma\,\sqrt{h}h^{ab}\,\frac{1}{Z^2}
 \left(\partial _aX^\mu \partial _bX_\mu +\partial _aZ\partial _bZ\right)+\ldots  
\end{equation}

The dependence on the shape of the loop originates from the boundary conditions for the string embedding coordinates:  the string worldsheet should end on the contour $C$ on the boundary of AdS (at $z=0$ in the Poincar\'e parameterization   (\ref{ZAPoincare})) and on the contour $n^I$ on $S^5$:
\begin{equation}
 X^\mu (\sigma ,0)=x^\mu (\sigma ),
 \qquad 
 Z(\sigma ,0)=0,
 \qquad 
 N^I(\sigma ,0)=n^I(\sigma ).
\end{equation}
At $\lambda \rightarrow \infty $ the string path integral is saturated by a saddle point and the Wilson loop expectation value obeys the minimal area law in $AdS_5\times S^5$. 

\begin{figure}[t]
\begin{center}
 \centerline{\includegraphics[width=8cm]{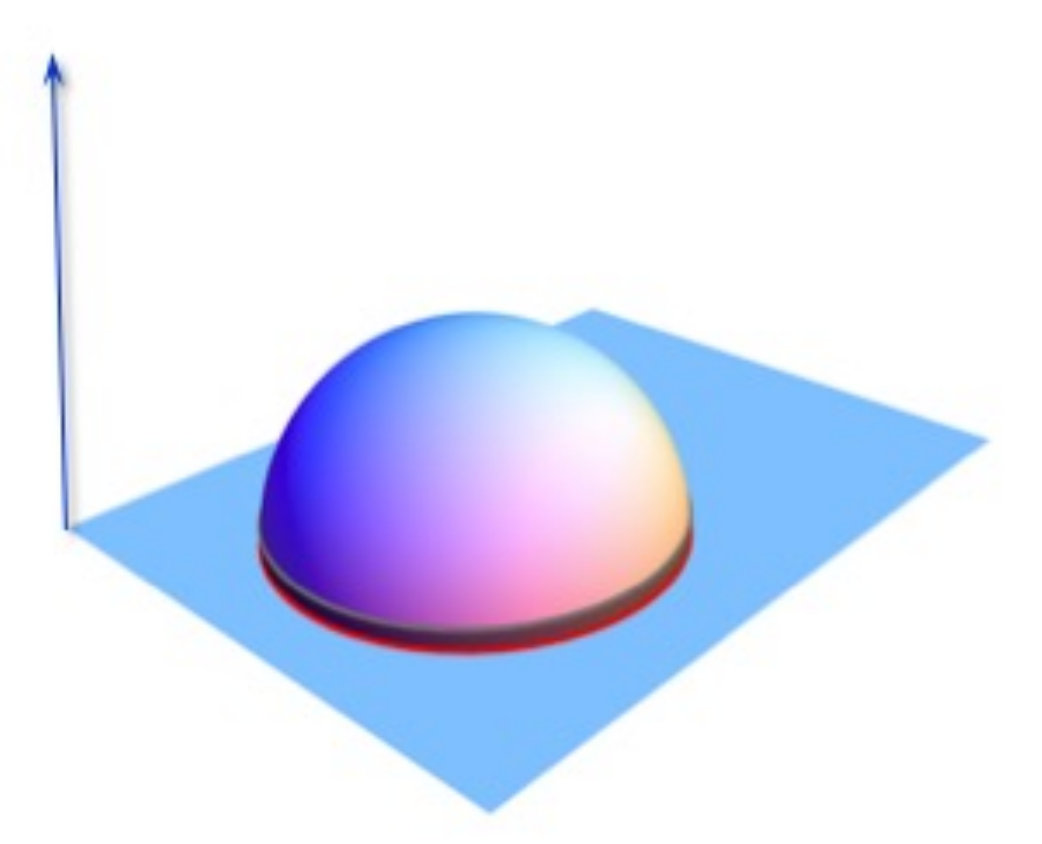}}
\caption{\label{ZAWci}\small The minimal surface for the circular Wilson loop. Its area is regularized by cutting out a boundary layer of thickness $\varepsilon $.}
\end{center}
\end{figure}

We thus need to find the minimal surface in the Anti-de-Sitter space that ends on a circle on the boundary. The solution can actually be obtained without solving any equations, just by exploiting the symmetries of the problem. For the straight line $x^\mu(\sigma ) =n_1^\mu/(2R) +\sigma n_2^\mu $, where $n_1^\mu $ and $n_2^\mu $ are two orthogonal unit vectors, the minimal surface is obvious:
\begin{equation}\label{ZAAdS2}
 X^\mu =\frac{n_1^\mu}{2R} +\sigma n_2^\mu ,\qquad Z=\tau ,
\end{equation}
which is an $AdS_2$ hyperplane embedded in $AdS_5$. The metric of $AdS_5$ admits the following isometry, which on the boundary
reduces to the inversion
 accompanied by a translation:
\begin{equation}
 X^\mu \rightarrow \frac{X^\mu }{Z^2+X^2}-Rn_1^\mu \,\qquad Z\rightarrow \frac{Z}{Z^2+X^2}\,.
\end{equation}
This transformation leaves the string action invariant, and  consequently maps solutions of the equations of motion to solutions. Applying this map to (\ref{ZAAdS2}), and changing the worldsheet coordinates as
\begin{equation}
 \frac{\tau }{R}\,\,\frac{1}{\frac{1}{4R^2}+\tau ^2+\sigma ^2}\rightarrow \tanh\tau ,
 \qquad 
 \frac{\sigma }{R}\,\,\frac{1}{\frac{1}{4R^2}-\tau ^2-\sigma ^2}\rightarrow \tan\sigma ,
\end{equation}
we arrive at the minimal surface for the circle  \cite{ZABerenstein:1998ij,ZADrukker:1999zq}:
\begin{equation}\label{ZAcirclems}
 X^\mu =\frac{R}{\cosh\tau }(n_1^\mu \cos\sigma +n_2^\mu \sin\sigma ),
 \qquad Z=R\tanh\tau .
\end{equation}
Geometrically this solution represents a hemisphere $X^2+Z^2=R^2$ in the bulk of AdS (fig.~\ref{ZAWci}).

The solution for the circle, written as (\ref{ZAcirclems}), is already in the conformal gauge, and to compute the area one can just plug it in the string action with $h_{ab}=\delta _{ab}$. The $\tau $ integral  then diverges because of the $1/Z^2$ factor in the AdS metric and has to be regularized. The correct renormalization prescription, justified in \cite{ZADrukker:1999zq}, is to cut off a boundary layer $Z<\varepsilon $ (as illustrated in fig.~\ref{ZAWci}), compute the regularized area, and apply an operator $1+\varepsilon \,\partial /\partial \varepsilon $ to the result. The  sole function of the last step is to remove the singular $1/\varepsilon $ term. After that one can send $\varepsilon $ to zero. Applying this procedure to the minimal surface (\ref{ZAcirclems}), we find:
\begin{equation}
 S_{\rm str,ren}(C_{\rm circle})=2\pi 
 \lim_{\varepsilon \rightarrow 0} 
 \left(1+\varepsilon \,\frac{\partial }{\partial \varepsilon }\right)
 \int_{\varepsilon }^{\infty }\frac{d\tau }{\sinh^2\tau }=-2\pi ,
\end{equation}

For the Wilson loop expectation value we thus get:
\begin{equation}\label{ZAWcr}
 W(C_{\rm circle})\simeq \,{\rm e}\,^{\sqrt{\lambda }},
\end{equation}
in complete agreement with the exact result (\ref{ZAStronglambdacircle}) \cite{ZAErickson:2000af}. The prefactor in the asymptotic expansion of the exact formula should come from string fluctuations around the minimal surface (\ref{ZAcirclems}). The factor of $\lambda ^{-3/4}$ was interpreted in \cite{ZADrukker:2000rr} as a leftover of the residual $SL(2,\mathbb{R})$ symmetry of the disc partition function. Each of the $SL(2,\mathbb{R})$ generators is accompanied by a factor of $\hbar^{1/2}\sim \lambda ^{-1/4}$ giving in total $\lambda ^{-3/4}$. The numerical constant, apart of the fluctuation determinants computed in \cite{ZAKruczenski:2008zk}, depends on the structure and normalization of the measure in the string path integral \cite{ZAKristjansen:2012nz}, a delicate issue that has not been sorted out yet (see \cite{ZABergamin:2015vxa,ZAForini:2015bgo,ZAFaraggi:2016ekd} for a recent discussion). 

On the matrix model side, the leading exponential behavior of the Wilson loop is dictated by the largest eigenvalue:
\begin{equation}\label{ZAexpWLoop}
 W(C_{\rm circle})\simeq \,{\rm e}\,^{2\pi \mu },
\end{equation}
because the measure in the integral (\ref{ZAintMMWL}) is exponentially peaked at the rightmost edge of the eigenvalue density. The strong-coupling expansion of the Wilson loop can be systematically constructed by expanding the eigenvalue density in power series in $\mu -x$. The integral representation (\ref{ZAintMMWL}) can then be regarded as the Borel transform of perturbation series in $1/\sqrt{\lambda }$, with the Borel variable $t=\sqrt{\lambda }-2\pi x$. The square-root branch cut at the smallest eigenvalue, corresponding to $t=2\sqrt{\lambda } $, renders the strong-coupling expansion non-Borel-summable. Interestingly, it has an instanton interpretation. There is an unstable solution of the string sigma-model with the action equal to $+2\pi $ \cite{ZADrukker:2006ga}, which by the standard argument produces a singularity in the Borel plane at $t=2\sqrt{\lambda }$.

\section{Higher representations and D-branes}\label{ZAsec:HigherRepsD-branes}

The circular loop in the fundamental representation probes the AdS/CFT duality at the planar level or, in the string-theory language, at the leading order in the string coupling. It is possible to access all orders in $1/N$  while still remaining in the realm of classical gravity  by considering Wilson loops in large representations whose rank scales with $N$: $k\sim N$. The fundamental string that ends on the Wilson line then puffs into a D-brane \cite{ZADrukker:2005kx}, which behaves classically at large $N$ and large $\lambda $. On the field theory side, the regime of $k\sim N$ requires resummation of all terms in the $1/N$ expansion enhanced by powers of $k$.

We concentrate on completely symmetric and completely anti-sym\-met\-ric representations (in the latter case $k$ is bounded $N$; for symmetric representations $k$ is arbitrary; more general representations are discussed in \cite{ZAOkuda:2008px,ZAFiol:2013hna}):
\begin{equation}
\ytableausetup{mathmode, smalltableaux, centertableaux}
 \mathcal{R}_k^+=
 \overbrace{
 \begin{ytableau}
 ~ & ~ & ~ & \none & \none[\dots] & \none  &  \\
 \end{ytableau}}^{k}
 \qquad 
 \mathcal{R}_k^-=
 \left.
 \begin{ytableau}
 ~ \\ ~ \\ ~ \\ \none \\ \none[\vdots] \\ \none \\ ~ \\
 \end{ytableau}
 \right\}k.
\end{equation}
Wilson loops in these representations depend on two variables in the large-$N$ limit:
\begin{equation}
 W_\pm\left(\lambda ,\frac{k}{N}\right)=W_{\mathcal{R}_k^\pm}(C_{\rm circle}).
\end{equation}

The characters of symmetric and anti-symmetric representations are conveniently packaged into the generating functions
\begin{equation}
 \chi _\pm(\nu,\Phi  )=\sum_{k}^{}\,{\rm e}\,^{-2\pi k\nu }\mathop{\mathrm{tr}}\nolimits_{\mathcal{R}^\pm_k}\,{\rm e}\,^{2\pi \Phi }.
\end{equation}
When expressed through eigenvalues, the generating functions of symmetric/anti-symmetric characters are equivalent to Bose or Fermi distributions:
\begin{equation}
 \chi _\pm(\nu ,\Phi )=\prod_{i=1}^{N}\left[1\mp\,{\rm e}\,^{2\pi \left(a_i-\nu \right)}\right]^{\mp 1}.
\end{equation}
Consequently,
\begin{equation}\label{ZAWpm}
 W_\pm(\lambda ,f)=-i\left\langle \int_{\Lambda -\frac{i}{2}}^{\Lambda +\frac{i}{2}}
d\nu \,\,{\rm e}\,^{2\pi Nf\nu } 
\prod_{i=1}^{N}\left[1\mp\,{\rm e}\,^{2\pi \left(a_i-\nu \right)}\right]^{\mp 1}
 \right\rangle,
\end{equation}
where $\Lambda $ is some large number, bigger than any $a_i$. 

In is convenient to think of eigenvalues  as (random) energy levels, $\nu $ then plays the r\^ole of the chemical potential and $f$ has the meaning of the particle density. The last formula then relates canonical and grand canonical partition functions of an $N$-level system of $k$ non-interacting particles.

The Bose/Fermi partition functions in (\ref{ZAWpm}) are exponentially large, but not as large as the action in the matrix integral -- the exponent is $O(N)$ compared to the $O(N^2)$ action. The insertion of the Wilson loop with $k\sim N$ therefore does not backreact on the saddle point of the matrix model, and the
average over the ensemble of random eigenvalues can be replaced, at large $N$, by average over the Wigner distribution (\ref{ZAWigner}). The integration over $\nu $  is also  saturated by a saddle point, and we get \cite{ZAHartnoll:2006is}:
\begin{equation}\label{ZAfreeEdef}
 W_\pm(\lambda ,f)\simeq \,{\rm e}\,^{2\pi NF_\pm(\lambda ,\nu )},
\end{equation}
where the free energy is given by
\begin{equation}\label{ZAfree-en}
 F_\pm(\lambda ,\nu )=f\nu \mp\frac{1}{2\pi }\int_{-\mu }^{\mu }dx\,\rho (x)\ln\left(1\mp\,{\rm e}\,^{2\pi \left(x-\nu \right)}\right).
\end{equation}
The chemical potential $\nu $ is determined by minimizing the free energy:
\begin{equation}\label{ZAf-dens}
 0=\frac{\partial F_\pm}{\partial \nu }=f-\int_{-\mu }^{\mu }\frac{dx\,\rho (x)}{\,{\rm e}\,^{2\pi \left(\nu -x\right)}\mp 1}\,.
\end{equation}

These are the standard textbook formulas for the partition function of a non-interacting Bose/Fermi gas with the single-particle level density $\rho (x)$.
We are mostly interested in the strong-coupling regime, when the effective temperature (of order one) is much smaller than the typical "energy" (that is, typical eigenvalue), which at strong coupling scales as $\sqrt{\lambda }$. Symmetric and anti-symmetric representations behave very differently in the low-temperature regime, and will be considered separately.

We begin with the anti-symmetric case. At low temperature the Fermi distribution is well approximated by the step function, and eq.~(\ref{ZAf-dens})  simplifies to
\begin{equation}
 f=\int_{\nu  }^{\mu }dx\,\rho (x),
\end{equation}
which for the Wigner density (\ref{ZAWigner}) gives:
\begin{equation}\label{ZApifteta}
 \pi f=\theta -\frac{1}{2}\sin 2\theta ,
\end{equation}
with
\begin{equation}
 \cos\theta =\frac{\nu }{\mu }\,.
\end{equation}
For the free energy we get, in the same approximation \cite{ZAYamaguchi:2006tq,ZAHartnoll:2006is}:
\begin{equation}\label{ZAantissymfree}
 F_-=\int_{\nu  }^{\mu }dx\,\rho (x)x=\frac{\sqrt{\lambda }}{3\pi ^2}\,\sin^3\theta .
\end{equation}
The standard low-temperature expansion of the Fermi-gas partition function generates the strong-coupling expansion of the Wilson loop. Explicit results for higher orders in $1/\sqrt{\lambda }$ can be found in \cite{ZAHorikoshi:2016hds}.

When the chemical potential $\nu $ changes between $+\mu $ to $-\mu $, the density $f$ increases from $0$ to $1$. This is consistent with the fact that anti-symmetric representations only exist for $k<N$, and so $f$ cannot exceed $1$. Moreover, representations with $k$ and $N-k$ boxes are complex conjugate to one another. In the above formulas the conjugation symmetry acts as $\theta \rightarrow \pi -\theta $ and leaves all the equations invariant as it should.

The symmetric case is more subtle, because the chemical potential for bosons must be negative, which in our conventions means that $\nu >\mu $. But as $\nu $ decreases from infinity to $\mu $, $f$ according to eq.~(\ref{ZAf-dens}) grows from zero to a finite value
\begin{equation}
 f_c=\frac{1}{2\pi }\int_{-\mu }^{\mu }\frac{dx\,\rho (x)}{\mu -x}=\frac{2}{\sqrt{\lambda }}\,,
\end{equation}
which moreover becomes very small at strong coupling. At larger densities equation (\ref{ZAf-dens}) has no solutions. In the statistical mechanics analogy this corresponds to the Bose-Einstein condensation. The Wilson loop expectation value,  however, does not have any thermodynamic singularity at $f=f_c$, and can be analytically continued past the critical point \cite{ZAHartnoll:2006is,ZAGrignani:2009ua}. 

In the Bose-Einstein condensed phase a contribution of the largest eigenvalue to the average (\ref{ZAWpm}) is macroscopically large, allowing for $f>f_c$ in thermodynamic equilibrium.  Alternatively one can compute the Wilson loop at $f<f_c$ and analytically continue the result past the critical point \cite{ZAHartnoll:2006is}.
The equivalence of the two prescriptions is not obvious, but can be proven under very general assumptions \cite{ZAChen-Lin:2015dfa}. We follow the derivation based on the  condensation picture.

\begin{figure}[t]
\begin{center}
 \centerline{\includegraphics[width=8cm]{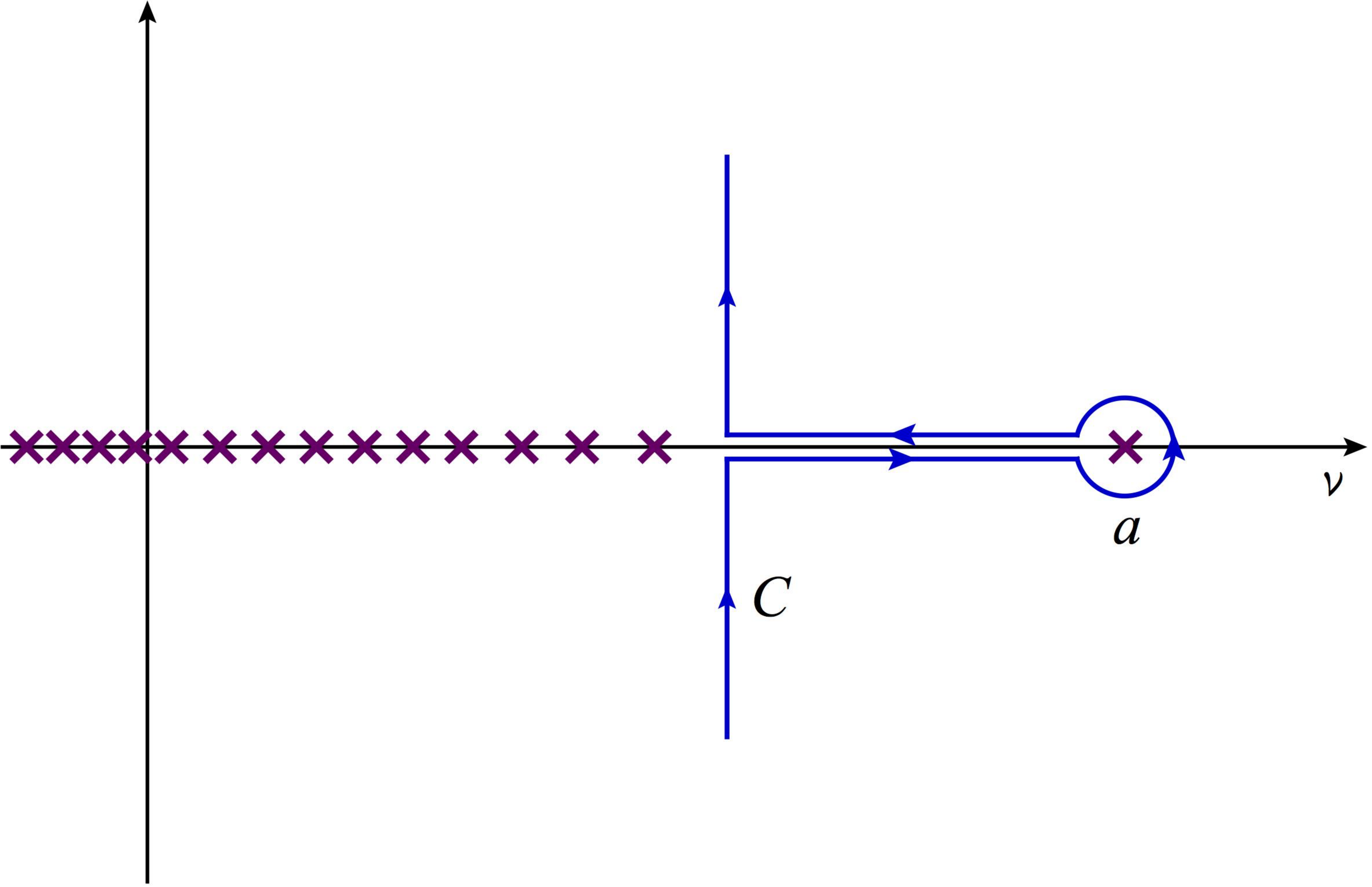}}
\caption{\label{ZAcoint}\small The contour of integration that singles out the contribution of the largest eigenvalue (from \cite{ZAChen-Lin:2015dfa}).}
\end{center}
\end{figure}

A contribution of the largest eigenvalue is singled out by contour deformation  in (\ref{ZAWpm}) shown in fig.~\ref{ZAcoint}. The integral then picks a residue at the largest of $a_i$'s, which we denote simply by $a$. It is important to emphasize that $a$ is actually different from $\mu $. There is a non-zero, albeit small probability to find an eigenvalue outside of the interval on which the macroscopic density is defined.  The smallness of this probability is counterbalanced by the exponentially large statistical weight in (\ref{ZAWpm}). 
Taking into account the extra price of pulling out an eigenvalue out of the macroscopic distribution, we get for the leading pole term in (\ref{ZAWpm}):
\begin{equation}\label{ZAintermsymmr}
 W_\pm=\int_{\mu }^{\infty }da\,P(a)\,{\rm e}\,^{2\pi Nfa}
 \prod_{i}^{}\frac{1}{1-\,{\rm e}\,^{2\pi \left(a_i-a\right)}}\,
\end{equation}
where $P(a)$ is the probability to find the largest eigenvalue at $a$. The latter can be read off from the partition function (\ref{ZAeigenvalueint}):
\begin{equation}
 P(a)=\,{\rm const}\,\,{\rm e}\,^{-\frac{8\pi ^2N}{\lambda }\,a^2}\prod_{i}^{}\left(a-a_i\right)^2.
\end{equation}
The normalization constant is determined by the condition that $P(\mu )=O(1)$, so the exponent should vanish at $a=\mu $. 

Evaluating the integral (\ref{ZAintermsymmr}) in the saddle-point approximation we find for the free energy defined in (\ref{ZAfreeEdef}):
\begin{equation}
 F_+=fa-\frac{4\pi }{\lambda }\,a^2+\frac{1}{2\pi }\int_{-\mu }^{\mu }
 dx\,\rho (x)\ln\frac{\left(a-x\right)^2}{1-\,{\rm e}\,^{2\pi \left(x-a\right)}}+F_0,
\end{equation}
where
\begin{equation}
 F_0=\frac{4\pi }{\lambda }\mu ^2-\frac{1}{\pi }\int_{-\mu }^{\mu }
dx\,\rho (x)\ln\left(\mu -x\right)=\frac{1}{2\pi }\,\ln\frac{16\pi ^2{\rm e}}{\lambda }\,.
\end{equation}
The  largest eigenvalue $a$ is determined by the saddle-point equation $\partial F_+/\partial a=0$:
\begin{equation}
 f=\frac{8\pi a}{\lambda }+\int_{-\mu }^{\mu }dx\,\rho (x)\left[
 \frac{1}{\,{\rm e}\,^{2\pi \left(a-x\right)}-1}-\frac{1}{\pi }\,\,\frac{1}{a-x}
 \right].
\end{equation}

At strong coupling when $a-\mu \sim \sqrt{\lambda }$, the first term under the integral is exponentially small in $\sqrt{\lambda }$ and can be neglected, which gives:
\begin{equation}
 f\simeq \frac{8\pi a}{\lambda }-\frac{1}{\pi }\int_{-\mu }^{\mu }\frac{dx\,\rho (x)}{a-x}=\frac{2}{\pi \mu ^2}\,\sqrt{a^2-\mu ^2}
\end{equation}
Introducing the rescaled variable
\begin{equation}\label{ZAnewkappa}
 \kappa =\frac{\sqrt{\lambda }f}{4}\,,
\end{equation}
we find that $a=\mu \sqrt{1+\kappa ^2}$ and \cite{ZADrukker:2005kx,ZAHartnoll:2006is}
\begin{equation}\label{ZAsymmetriccircle}
 \pi F_+=\kappa \sqrt{1+\kappa ^2}+\mathop{\mathrm{arcsinh}}\kappa .
\end{equation}

The string dual of a Wilson loop in the rank-$k$ representation is an object that carries $k$ units of the string charge. When $k$ is large, of order $N$, a natural candidate is a D-brane \cite{ZADrukker:2005kx,ZAYamaguchi:2006tq,ZAGomis:2006sb,ZARodriguezGomez:2006zz,ZAGomis:2006im,ZAYamaguchi:2007ps,ZAFiol:2014vqa}. This can be justified by considering brane intersections \cite{ZAGomis:2006sb,ZAGomis:2006im}, and relies on the following mechanism. Consider a Dp-brane whose $(p+1)$-dimensional worldvolume locally looks like $\Sigma \times S^{p-1}$, where $\Sigma $ is a two-dimensional surface that we identify with the string worldsheet. For the sake of the argument we may visualize $\Sigma $ as being "macroscopic", extending to large distances, while $S^{p-1}$ being "very small", such that from far apart the worldvolume appears two-dimensional. For the D-brane to carry the correct string charge it should couple to the $B_{MN}$ field as the fundamental string does.

The D-brane coupling to $B_{MN}$ arises from the DBI action:
\begin{equation}
 S_{\rm DBI}=T_{\rm Dp}\int_{}^{}d^{p+1}\sigma \,
 \sqrt{\det_{\mu \nu }\left(g_{\mu \nu }+B_{\mu \nu }+\frac{1}{T_{\rm F1}}\,F_{\mu \nu }\right)}\,,
\end{equation}
where $g_{\mu \nu }$ and $B_{\mu \nu }$ are pullbacks of the target-space fields, $F_{\mu \nu }$ is the internal gauge field on the D-brane worldvolume, $T_{{\rm Dp}}$ is the D-brane tension and $T_{\rm F1}$ is the tension  of the fundamental string. Expanding to the linear order in $B_{\mu \nu }$, we find:
\begin{equation}
 S_{{\rm DBI}}\ni T_{\rm F1}\int_{}^{}d^{p+1}\sigma \,B_{\mu \nu }\Pi ^{\mu \nu }+\ldots ,
\end{equation}
where
\begin{equation}
 \Pi ^{\mu \nu }=\frac{\delta S_{\rm DBI}}{\delta F_{\mu \nu }}\,.
\end{equation}

This should be compared to the string coupling to $B_{MN}$ (the coupling is pure imaginary if the worldsheet is Euclidean):
\begin{equation}
  S_{{\rm str}}\ni\frac{i}{2}\,T_{\rm F1}\int_{}^{}d^2\sigma \,B_{ab}\varepsilon ^{ab},
\end{equation}
where $B_{ab}$ is the pullback of $B_{MN}$ onto the worldsheet.
The D-brane will carry the correct amount $k$ of the string charge provided that the electric field $\Pi ^{\mu \nu }$ has components only along $\Sigma $, upon averaging over the sphere, and is normalized as
\begin{equation}
 \int_{S^{p-1}}^{}d^{p-1}\sigma \, \Pi ^{ab}=\frac{ik}{2}\,\varepsilon ^{ab}.
\end{equation}
This can be achieved by adding a Lagrange multiplier to the D-brane action:
\begin{equation}
 S_{\rm L. m.}=-\frac{ik}{2}\int_{\Sigma }^{}d^2\sigma \,\varepsilon ^{ab}F_{ab}=
 -ik\oint_C d\sigma ^a\,A_a.
\end{equation}
Here $C$ is the contour  on the boundary of $AdS_5$ at which the D-brane ends: $\partial \Sigma =C$. 

In the natural AdS units, in which the radius of $AdS_5$ is set to one, the D-brane tensions can be obtained from the standard formulas \cite{ZApolchinski1998string} by replacing $\alpha '\rightarrow 1/\sqrt{\lambda }$, $g_s\rightarrow  \lambda /4\pi N$:
\begin{equation}
 T_{\rm F1}=\frac{\sqrt{\lambda }}{2\pi }\,,
 \qquad 
 T_{{\rm D1}}=\frac{2N}{\sqrt{\lambda }}\,,
 \qquad 
 T_{\rm D3}=\frac{N}{2\pi ^2}\,,
 \qquad 
 T_{{\rm D5}}=\frac{N\sqrt{\lambda }}{8\pi ^4}\,.
\end{equation}
The D-brane tensions are factor of $N$ larger than the tension of the fundamental string. The D1-brane is dual to the 't~Hooft loop in the gauge theory, the D3-branes describe Wilson loops in the symmetric representations \cite{ZAGomis:2006im,ZAYamaguchi:2007ps}, and the D5-brane in the anti-symmetric \cite{ZAYamaguchi:2006tq,ZAGomis:2006sb}. The rank of the representation is determined by the electric flux on the D-brane world volume, as we have discussed above.

Consider first the D3-brane \cite{ZADrukker:2005kx}. Collecting together the DBI action, the Wess-Zumino coupling to the five-form potential and the Lagrange-multiplier term we get for the D3-brane action:
\begin{equation}\label{ZAD3S}
 S_{\rm D3}=\frac{N}{2\pi ^2}\int_{}^{}d^4x\,\left[
 \sqrt{\det_{\mu \nu }\left(g_{\mu \nu }+\frac{2\pi }{\sqrt{\lambda }}\,F_{\mu \nu }\right)}-\frac{1}{4!}\varepsilon ^{\mu \nu \lambda \rho }C_{\mu \nu \lambda \rho }
 \right]-ik\oint_C A
\end{equation}
As before, we will first solve the problem for the straight line and then get the result for the circle by a conformal transformation. 

The D3-brane dual to the straight line has an $AdS_2\times S^2$ shape, where $AdS_2$ is the original string worldsheet in the $(xz)$ plane,
and $S^2$ is the round sphere linking the $x$ axis in $\mathbb{R}^4$. The radius of the sphere evolves along the holographic direction, such that the D-brane embedding can be parameterized by $r=r(z)$. At the boundary of $AdS_5$ the D-brane should shrink to the Wilson line  so the boundary condition at $z=0$ is $r(0)=0$. The potential of the RR five-form that supports the $AdS_5\times S^5$ geometry, in a convenient gauge is given by
\begin{equation}
 C=\frac{r^2}{z^4}\,dx\wedge dr\wedge {\rm Vol}(S^2).
\end{equation}
Introducing the rescaled field strength,
\begin{equation}
 F\equiv \frac{2\pi }{\sqrt{\lambda }}\,F_{xz},
\end{equation}
we arrive at the reduced D-brane action:
\begin{equation}
 S_{\rm D3}=\frac{2N}{\pi }
 \int dx\,dz\,\left[
 \frac{r^2}{z^4}\left(\sqrt{\acute{r}^2+1+z^4F^2}-\acute{r}\right)
 -i\kappa F
 \right],
\end{equation}
where $\kappa $ is defined in (\ref{ZAnewkappa}).

The equations of motion that follow from this action are
\begin{eqnarray}
 && \frac{r^2F}{\sqrt{\acute{r}^2+1+z^4F^2}}=i\kappa 
\nonumber \\
&&\left[\frac{r^2}{z^4}\left(\frac{\acute{r}}{\sqrt{\acute{r}^2+1+z^4F^2}}-1\right)\right]'=
\frac{2r}{z^4}\left(\sqrt{\acute{r}^2+1+z^4F^2}-\acute{r}\right).
\end{eqnarray}
In spite of their complicated appearance, they have a simple solution:
\begin{equation}\label{ZADrukkerFiol}
 r=\kappa z,\qquad F=\frac{i}{z^2}\,.
\end{equation}
The action on this solution diverges, but cutting off the divergence at $z=\varepsilon $ and applying $1+\varepsilon \,\partial /\partial \varepsilon $ we get zero, in accord with  non-renormalization of the straight Wilson line, whose expectation value is $W_k({\rm line})=1$ as expected.

The solution for the circular loop can be obtained by inversion, but looks rather complicated in the standard Poincar\'e coordinates. The problem can be greatly simplified by a judicious choice of coordinates \cite{ZAYamaguchi:2006tq}. 
Applying a coordinate transformation
transformation $(r,z)\rightarrow (u,\zeta )$:
\begin{equation}
 r=\zeta  \tanh u,\qquad z=\frac{\zeta }{\cosh u},
\end{equation}
and substituting it into the $AdS_5$ metric
\begin{equation}
 ds^2_{AdS_5}=\frac{dz^2+dx^2+dr^2+r^2d\Omega _{S^2}^2}{z^2}\,,
\end{equation}
we get:
\begin{eqnarray}
 ds_{AdS_5}^2&=&du^2+\cosh^2u\,\frac{d\zeta ^2+dx^2}{\zeta ^2}+\sinh^2 u\,d\Omega ^2_{S^2}
\nonumber \\
&=&
 du^2+\cosh^2u\,d\Omega ^2_{AdS_2}+\sinh^2 u\,d\Omega ^2_{S^2}.
\end{eqnarray}
The boundary is now at $u\rightarrow \infty $ (or $\zeta \rightarrow 0$) and has the geometry of $AdS_2\times S^2$. This slicing of the AdS space may look unusual, but
there is no contradiction, since $\mathbb{R}^4$ is conformally equivalent to $AdS_2\times S^2$. This is easily seen by writing the Euclidean metric as
\begin{equation}
 ds_{\mathbb{R}^4}^2=r^2\left(\frac{dx^2+dr^2}{r^2}+d\Omega ^2_{S^2}\right).
\end{equation}
The solution (\ref{ZADrukkerFiol}) in the new coordinates is simply
\begin{equation}
 \sinh u=\kappa ,\qquad F=\frac{i\sqrt{1+\kappa ^2}}{\zeta ^2}\,.
\end{equation}
The D-brane sits at constant $u$, and the electric field is proportional to the volume form of $AdS_2$.

To obtain the solution for the circle we simply replace the Poincar\'e coordinates in $AdS_2$ by the global coordinates:
\begin{equation}
 ds_{AdS_5}^2=du^2+\cosh^2u\left(d\chi ^2+\sinh^2\chi \,d\varphi ^2\right)+\sinh^2 u\,d\Omega ^2_{S^2}.
\end{equation}
The boundary at $u=\infty $ is still $AdS_2\times S^2$ conformally equivalent to $\mathbb{R}^4$.  For the D-brane solution we again can take the hypersurface that spans $AdS_2\times S^2$ at constant $u$, with the electric field proportional to the volume form on $AdS_2$:
\begin{equation}
 \sinh u=\kappa ,\qquad F\equiv \frac{2\pi }{\sqrt{\lambda }}\,F_{\varphi \chi }=
 i\sqrt{1+\kappa ^2}\sinh\chi .
\end{equation}
The four-form potential in these coordinates is
\begin{equation}
 C=\frac{1}{2}\left(2\sinh^3u\cosh u+\sinh u\cosh u -u\right)\sinh\chi \,
 d\varphi \wedge d\chi \wedge {\rm Vol(S^2)}.
\end{equation}
Substituting the solution into the D3-brane action (\ref{ZAD3S}) we get:
\begin{equation}
 S_{\rm D3}=\frac{N}{2\pi^2 }\times \frac{1}{2}\,\left(\kappa \sqrt{1+\kappa ^2}+\mathop{\mathrm{arcsinh}}\kappa \right)\times {\rm Vol\left(S^2\right)}\times 
 \int_{0}^{2\pi }d\varphi \,\int_{0}^{\ln\frac{2}{\varepsilon }}d\chi \,\sinh\chi .
\end{equation}
The last, divergent factor is the volume of $AdS_2$, which as usual should be regularized by subtracting the $1/\varepsilon $ term. The renormalized volume of $AdS_2$ then equals $-2\pi $, and for the D-brane action we get \cite{ZADrukker:2005kx}:
\begin{equation}
 S_{{\rm D3, ren}}=-2N\left(\kappa \sqrt{1+\kappa ^2}+\mathop{\mathrm{arcsinh}}\kappa \right),
\end{equation}
in complete agreement with the matrix-model prediction (\ref{ZAfreeEdef}), (\ref{ZAsymmetriccircle}).

If $\kappa $ is small, 
\begin{equation}
 W_+\simeq \,{\rm e}\,^{4N\kappa }=\,{\rm e}\,^{\sqrt{\lambda }k}.
\end{equation}
This is just the $k$-th power of the Wilson loop in the fundamental representation (\ref{ZAWcr}). The character of a rank-$k$ representation can be expressed through ordinary traces and
for small $k\ll N$ only the term with the largest number of traces contributes  due to the large-$N$ factorization:
\begin{equation}
 \left\langle\mathop{\mathrm{tr}}\nolimits_{\mathcal{R}^\pm _k}\,{\rm e}\,^{2\pi \Phi }
 \right\rangle=
 \frac{1}{k!}\,\left\langle\left(\mathop{\mathrm{tr}}\,{\rm e}\,^{2\pi \Phi }\right)^k\right\rangle +\left\langle \mathcal{O}(\mathop{\mathrm{tr}}\nolimits^{k-1})
 \right\rangle
 =\frac{N^k}{k!}\,W_\Box^k+\mathcal{O}(N^{k-1}).
\end{equation}
For larger $\kappa $ the result starts to deviate from the simple $k$-th power of the fundamental  loop and consequently non-planar diagrams start to contribute. The complete result entails resummation of the   $(\lambda k^2/N^2)^n$ terms in the perturbative series and thus receives contributions from all orders of the $1/N$ expansion. This calculation therefore probes the AdS/CFT duality beyond the planar approximation.

A Wilson loop in an anti-fundamental representation is dual to a D5-brane. The classical solution in that case \cite{ZAYamaguchi:2006tq} is quite a bit simpler, because the D5-brane expands in $S^5$ rather than $AdS_5$. The expanded geometry has the direct product structure $\Sigma \times S^4$ not just locally but over the whole worldvolume of the D-brane. The four-sphere wraps a latitude on $S^5$ at a fixed polar angle $\theta $. The action of the D5-brane is
\begin{equation}
 S_{\rm D5}=\frac{N\sqrt{\lambda }}{8\pi ^4}\left[\int_{}^{}d^6x\,\sqrt{\det_{\mu \nu }\left(g_{\mu \nu }+\frac{2\pi }{\sqrt{\lambda }}\,F_{\mu \nu }\right)}-\frac{2\pi i}{\sqrt{\lambda }}\int F\wedge C \right]-ik\oint_C A.
\end{equation}
As before it is convenient to represent the last term as a volume integral over $\Sigma $. Using the product structure of the D-brane's worldvolume, one can  integrate by parts the Wess-Zumino term, and since $dC=-4{\rm Vol(S^5)}$, the integral of $C$ gives  four times the volume  enclosed by $S^4$ inside  $S^5$:
\begin{equation}
 \int_{S^4}^{}C=-4\times \frac{8\pi ^2}{3}\,\int_{0}^{\theta }d\psi \,\sin^4\psi =
- 4\pi ^2\left(\theta -\sin\theta \cos\theta -\frac{2}{3}\,\sin^3\theta \cos\theta \right).
\end{equation}
Parametrizing $\Sigma $ by the holographic coordinate $z$ and the polar angle $\varphi $ on the boundary, and absorbing the factor of $2\pi /\sqrt{\lambda }$ into $F_{\varphi z}$, we get:
\begin{equation}\label{ZASD5}
 S_{\rm D5}=\frac{N\sqrt{\lambda }}{\pi ^2}\int_{}^{}d\varphi \,dz\,\left[
 \frac{A}{z^2}\sqrt{r^2\left(\acute{r}^2+1\right)+z^4F^2}
 -iBF
 \right],
\end{equation}
where
\begin{eqnarray}
 A&=&\frac{1}{3}\,\sin^4\theta 
\nonumber \\
B&=&\frac{1}{2}\left(\pi f-\theta +\sin\theta \,\cos\theta +\frac{2}{3}\,\sin^3\theta \,\cos\theta \right).
\end{eqnarray}
Here $f=k/N$  and does not depend on $\sqrt{\lambda }$, in contradistinction to the D3-brane case. This is because a D5-brane is a factor of $\sqrt{\lambda }$ heavier than a D3-brane.

The equations of motion for (\ref{ZASD5}) are
\begin{eqnarray}
&& \frac{Az^2F}{\sqrt{r^2\left(\acute{r}^2+1\right)+z^4F^2}}=iB
\nonumber \\
&&\left(\frac{r^2}{z^2}\,\,\frac{\acute{r}}{\sqrt{r^2\left(\acute{r}^2+1\right)+z^4F^2}}\right)'
=
\frac{r}{z^2}\,\,\frac{\acute{r}^2+1}{\sqrt{r^2\left(\acute{r}^2+1\right)+z^4F^2}}
\end{eqnarray}
Their solution is  very simple -- the string worldsheet in AdS is undeformed:
\begin{equation}
 r=\sqrt{R^2-z^2}\,,
\end{equation}
and the field strength is equal to
\begin{equation}
 F=i\,\frac{B}{\sqrt{A^2+B^2}}\,\,\frac{R}{z^2}\,.
\end{equation}
Plugging this into the action (\ref{ZASD5}) we get:
\begin{equation}
 S_{\rm D5}=\frac{2N\sqrt{\lambda }}{\pi }\,\sqrt{A^2+B^2}\,R\int_{\varepsilon }^{R}\frac{dz}{z^2}\,.
\end{equation}
After subtracting the $1/\varepsilon $ divergence, this becomes
\begin{equation}
 S_{\rm D5, ren}=-\frac{2N\sqrt{\lambda }}{\pi }\,\sqrt{A^2+B^2}
\end{equation}

The position of the D5-brane on $S^5$ is determined by minimization of the on-shell action in $\theta $. Using the equality $\partial B/\partial \theta =-4A$, we find:
\begin{equation}
 \frac{\partial }{\partial \theta }\left(A^2+B^2\right)=2A\left(\frac{\partial A}{\partial \theta }-4B\right)=
 \frac{4}{3}\,\sin^4\theta\left(\theta -\sin\theta \,\cos\theta -\pi f\right). 
\end{equation}
Consequently,
\begin{equation}
 \pi f=\theta -\frac{1}{2}\,\sin 2\theta .
\end{equation}
For the renormalized action we thus get:
\begin{equation}
 S_{\rm D5, ren}=-\frac{2N\sqrt{\lambda }}{3\pi }\,\sin^3\theta .
\end{equation}
Again this is in perfect agreement \cite{ZAYamaguchi:2006tq} with the matrix-model predictions (\ref{ZAfreeEdef}), (\ref{ZApifteta}) and (\ref{ZAantissymfree}). 

The holographic calculations described above are purely classical, which is justified by a combination of the large-$N$ and strong-coupling limits. Since the D-brane tension is proportional to $N$, quantum fluctuations  are $1/N$ suppressed. The one-loop fluctuation corrections have been actually calculated for the D5-brane in   \cite{ZAFaraggi:2011ge,ZAFaraggi:2014tna} and for the D3-brane in  \cite{ZABuchbinder:2014nia,ZAFaraggi:2014tna}, but the results so far disagree with the $1/N$ corrections in the matrix model, which can also be accounted for.
The disagreement does not necessarily mean that the relationship between Wilson loops  and D-branes only holds at the leading order. It may well be that the $1/N$ expansion on the matrix-model side contains some subtle contributions which have been overlooked, or the backreaction of the D-branes on the geometry (which for a single D-brane is a $1/N$ effect) has not been properly taken into account. It would be very interesting to resolve this apparent contradiction.

\section{Wavy lines, cusp and latitude}\label{ZAsec:Wavy}

In this section we consider,  following \cite{ZACorrea:2012at}, a number of Wilson loop observables -- the wavy lines, the cusp anomalous dimension, the heavy quark potential and the circular  latitude on $S^5$.  At first sight they seem unrelated but in fact can be expressed through one another, and some of them can be computed with the help of localization.

\begin{figure}[t]
\begin{center}
 \centerline{\includegraphics[width=10cm]{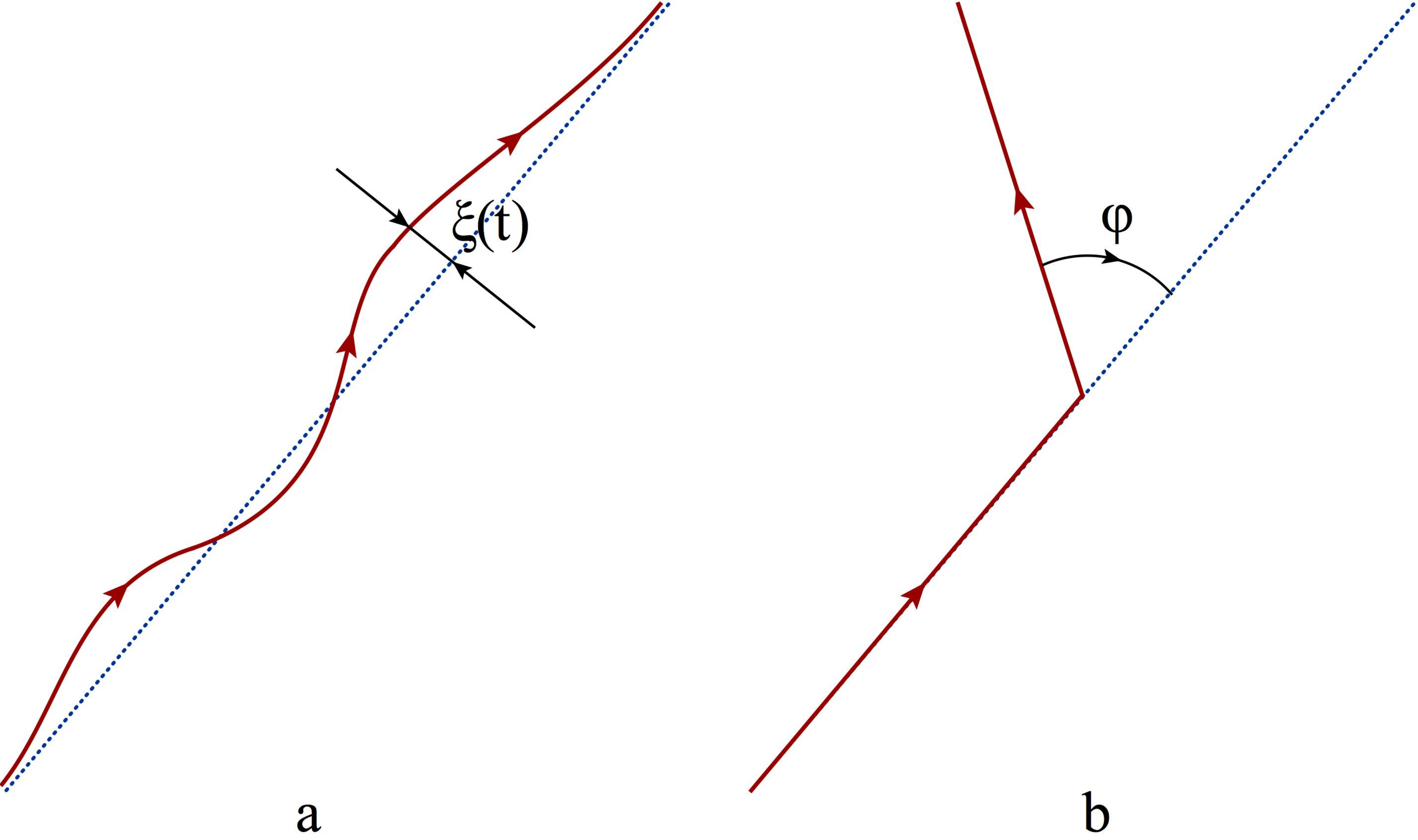}}
\caption{\label{ZAwavies}\small (a) A wavy line. (b) Cusp with an opening angle $\pi -\phi $.}
\end{center}
\end{figure}

A wavy line \cite{ZAPolyakov:2000ti,ZASemenoff:2004qr} (fig.~\ref{ZAwavies}a) is the straight line with a small perturbation on top: $x^\mu (t)=\delta ^\mu _0t+\xi ^\mu (t)$. Its expectation value  starts at the quadratic order in waviness. The functional form of the leading piece is completely fixed by translational invariance and scale symmetry \cite{ZASemenoff:2004qr}:
\begin{equation}\label{ZAbrems}
 \frac{1}{N}\,W(C_\xi )-1=\frac{B}{2}\int_{-\infty }^{+\infty }dt_1\,dt_2\,\,\frac{\left(\dot{\xi }(t_1)-\dot{\xi }(t_2)\right)^2}{\left(t_1-t_2\right)^2}+\mathcal{O}\left(\xi ^4\right).
\end{equation}
The coefficient $B\equiv B(\lambda )$ is  called  the Bremsstrahlung function \cite{ZACorrea:2012at}. 

It is instructive to see how this structure arises at the leading order in perturbation theory. Expanding the sum of scalar and vector propagators, eq.~(\ref{ZA1-loop}), in $\xi $ we get:
\begin{equation}
  \frac{|1+\dot{\xi }_1|\,|1+\dot{\xi }_2|-(1+\dot{\xi }_1\cdot \dot{\xi }_2)}{\left(t_1-t_2\right)^2+\left(\xi _1-\xi _2\right)^2}=
  \frac{\frac{1}{2}\left(\dot{\xi }(t_1)-\dot{\xi }(t_2)\right)^2}{\left(t_1-t_2\right)^2}
  +\mathcal{O}\left(\xi ^4\right),
\end{equation}
which gives for the weak-coupling asymptotics of the Bremsstrahlung function \cite{ZASemenoff:2004qr}:
\begin{equation}
 B(\lambda )\simeq \frac{\lambda }{16\pi ^2}\qquad \left(\lambda \rightarrow 0\right).
\end{equation}
In QED, where only the vector exchange is present, the wavy line could be brought to the same form by subtracting the linearly divergent self-energy. The Bremsstrahlung function then is $B_{\rm QED}= 2\alpha /3\pi $ at the leading order of perturbation theory.

Another quantity of interest is the cusp anomalous dimension. If a contour has a cusp, the associated Wilson loop develops a logarithmic singularity due to locally divergent diagrams \cite{ZAPolyakov:1980ca}. 
 Since the anomaly is a local effect, it can be studied by zooming onto the  vicinity of the cusp and considering the contour shown in fig.~\ref{ZAwavies}b. The expectation value of an infinite cusp diverges both in the UV and in the IR, and
needs regularization. The natural IR cutoff is the scale $L$ at which the Wilson loop starts to deviate from the simple straight-line cusp. To implement the UV cutoff one can round off the tip of the cusp on the scale of order $\varepsilon \ll L$. 

The expectation of a Wilson loop with a cusp behaves as
\begin{equation}
W(C_{\rm cusp})=\,{\rm const}\,\left(\frac{L}{\varepsilon }\right)^{\Gamma _{\rm cusp}(\phi ,\lambda )}.
\end{equation}
The exponent $\Gamma _{\rm cusp}(\varphi ,\lambda )$ is called the cusp anomalous dimension and depends on the opening angle of the cusp  and  the 't~Hooft coupling. It can be computed order by order in perturbation theory.
The cusp anomalous dimension has important applications in QCD \cite{ZAKorchemsky:1992xv}, and has been extensively studied in the context of the AdS/CFT duality \cite{ZADrukker:1999zq,ZAMakeenko:2006ds,ZADrukker:2011za,ZACorrea:2012nk,ZAHenn:2013wfa}.

At the first order of perturbation theory, we get from (\ref{ZA1-loop}):
\begin{eqnarray}
 \frac{1}{N}\,W(C_{\rm cusp})&=&1+\frac{\lambda }{8\pi ^2}
 \int_{\varepsilon }^{L}ds\,dt\,\,\frac{1-\cos\phi }{s^2+t^2+2st\cos\phi }
\nonumber \\
&=&1+\frac{\lambda }{8\pi ^2 }\,\phi \,\frac{1-\cos\phi }{\sin\phi }\int_{\varepsilon }^{L}\frac{dt}{t}+{\rm finite}.
\end{eqnarray}
The integral over $t$ produces the divergent logarithm, 
and for the one-loop cusp anomaly we obtain \cite{ZADrukker:1999zq}:
\begin{equation}\label{ZAcusp1loop}
 \Gamma _{\rm cusp}(\phi ,\lambda )=\frac{\lambda }{8\pi ^2 }\,\phi \tan\frac{\phi }{2}
 \qquad \left(\lambda \rightarrow 0\right).
\end{equation}

This formula has a number of interesting limits. It can be analytically continued to pure imaginary angles: $\phi \rightarrow i\theta  $, which is equivalent to changing the Euclidean cusp into a contour in the Minkowski space. The cusp then corresponds to a trajectory of a real particle that experiences an instantaneous acceleration. As can be seen from the leading-order result (\ref{ZAcusp1loop}), but is true more generally, the cusp anomaly is a growing function of rapidity with linear asymptotics:
\begin{equation}
 \Gamma _{\rm cusp}(i\theta ,\lambda )=-4f(\lambda )\theta \qquad 
 \left(\theta \rightarrow \infty \right).
\end{equation}

The function $f(\lambda )$, that characterizes the light-like cusp, is also referred to as the cusp anomalous dimension. It is  related to the scaling dimensions of twist-2 local operators:
\begin{equation}
 \mathcal{O}_S=\mathop{\mathrm{tr}}ZD_+^SZ,
\end{equation}
where $Z=\Phi _1+i\Phi _2$ and $D=D_1+D_2$. The twist-2 anomalous dimension grows logarithmically with the spin, and the coefficient coincides with the cusp anomaly:
\begin{equation}
 \gamma _S(\lambda )\simeq f(\lambda )\ln S\qquad \left(S\rightarrow \infty \right).
\end{equation}

When the opening angle approaches $\pi $ (equivalently, $\phi \rightarrow 0$), the cusp becomes the straight line, whose expectation values is finite, and is actually trivial in $\mathcal{N}=4$ SYM. The anomalous dimension should consequently vanish  at $\phi=0 $. Its Taylor expansion starts at the second order and is expressed through the Bremsstrahlung function. The explicit one-loop result (\ref{ZAcusp1loop}) is in accord with these observations.  Indeed, the cusp with a very small deflection angle can be viewed as a particular case of the wavy line, and its expectation value can be thus extracted from the general formula (\ref{ZAbrems}) by substituting $\dot{\xi }^\mu =\theta (t)\phi n^\mu $, where $n^\mu $ is the unit normal to the first segment of the cusp:
\begin{equation}
 \frac{1}{N}\,W(C_{\rm cusp})-1\stackrel{\phi \rightarrow 0}{=}
 B\phi ^2\int_{\varepsilon }^{L}\frac{dt_1\,dt_2}{\left(t_1+t_2\right)^2}=
 B\phi ^2\ln\frac{L}{\varepsilon }\,,
\end{equation}
which implies that
\begin{equation}\label{ZAbremfromcusp}
 \Gamma _{\rm cusp}(\phi ,\lambda )\stackrel{\phi \rightarrow 0}{=}B(\lambda )\phi ^2 .
\end{equation}

Finally, the quark-anti-quark potential can be also expressed through the cusp anomaly. Because of the conformal invariance, the potential in $\mathcal{N}=4$ SYM obeys the Coulomb law:
\begin{equation}
 V(L,\lambda )=-\frac{\alpha (\lambda )}{L}\,,
\end{equation}
and is characterized by a single functions of the 't~Hooft coupling, the Coulomb charge $\alpha $. 

Normally, the quark-anti-quark potential is associated with the long rectangular contour, but it can also be extracted from the cusp anomaly.
The cusped Wilson loop physically corresponds to a quark-anti-quark pair created at the tip of the cusp, the two particles flying  apart at constant velocity. When the opening angle of the cusp is very small ($\phi \rightarrow \pi $), the relative velocity is also small and the interaction between the particles is dominated by the quasi-static Coulomb energy:
\begin{equation}
 \ln W(C_{\rm cusp})\simeq \int_{\varepsilon }^{L} dt\,\,\frac{\alpha }{2t\sin\frac{\pi -\phi }{2} }\simeq \frac{\alpha }{\pi -\phi }\ln\frac{L}{\varepsilon }\,.
\end{equation}
Consequently,
\begin{equation}\label{ZApotfromcusp}
 \alpha (\lambda )=\lim_{\phi \rightarrow \pi }\left(\pi-\phi  \right)\Gamma _{\rm cusp}(\phi ,\lambda ).
\end{equation}
In particular, at weak coupling we get:
\begin{equation}
 \alpha (\lambda)=\frac{\lambda }{4\pi }\qquad \left(\lambda \rightarrow 0\right).
\end{equation}

At strong coupling the cusp anomalous dimension is determined by the area of the minimal surface in $AdS_5$ ending on the cusp at the boundary \cite{ZADrukker:1999zq}. Due to the symmetries of the problem, the solution has a self-similar form. In  the polar coordinates $(r,\varphi )$ centred at the tip of the cusp the minimal surface can be parameterized as
\begin{equation}
 z=ru(\varphi ).
\end{equation}
The Nambu-Goto action evaluated on this ansatz is
\begin{equation}\label{ZAcusp-sol}
 S_{\rm str}=\frac{\sqrt{\lambda }}{2\pi }\int_{}^{}\frac{dr}{r}\,\,d\varphi \,\,
 \frac{1}{u^2}\,\sqrt{1+u^2+\acute{u}^2}.
\end{equation}
Integration over $r$ diverges logarithmically and gives the requisite $\ln(L/\varepsilon )$ factor.

The equations of motion for $u$ admit a first integral, due to translational symmetry in the angular direction, which can be used to solve for $\acute{u}$:
\begin{equation}\label{ZAacuteu}
 \acute{u}=\frac{1}{u^2}\,\sqrt{\left(u_0^2-u^2\right)\left(u^2+1\right)
 \left(u^2+\frac{u_0^2}{1+u_0^2}\right)}\,,
\end{equation}
where $u_0$ is the constant of integration. Geometrically, $u_0$ is the maximum of $u(\varphi )$, which due to the symmetries of the problem is reached at $\varphi =(\pi -\phi )/2$. Consequently,
\begin{equation}
 \frac{\pi -\phi }{2}=\int_{0}^{u_0}\frac{du}{\acute{u}}\,.
\end{equation}
The integration yields:
\begin{equation}\label{ZAimplangle}
 \frac{\pi -\phi }{2}=\frac{1}{u_0}\,\sqrt{\frac{1+u_0^2}{2+u_0^2}}
 \left[
 \left(1+u_0^2\right)\Pi (-u_0^2)-K
 \right],
\end{equation}
where $\Pi (n)\equiv \Pi (n|m)$ and $K\equiv K(m)$ are the standard elliptic integrals of the third and first kind with the modulus given by
\begin{equation}
 m=\frac{1}{2+u_0^2}\,.
\end{equation}
Changing the integration variable in (\ref{ZAcusp-sol}) from $\varphi $ to $u$ with the help of (\ref{ZAacuteu}), and  subtracting the usual $1/\varepsilon $ divergence near the boundary, we get for the cusp anomaly at strong coupling \cite{ZADrukker:1999zq,ZADrukker:2007qr}:
\begin{equation}
 \Gamma _{\rm cusp}=\frac{\sqrt{\lambda }}{\pi u_0\sqrt{2+u_0^2}}
 \left[\left(2+u_0^2\right)E-\left(1+u_0^2\right)K\right],
\end{equation}
where $u_0$ is expressed through $\phi $ by inverting (\ref{ZAimplangle}).

The strong-coupling behavior of the Coulomb charge  can be extracted from the above formulas by taking the $u_0\rightarrow 0$ limit and using (\ref{ZApotfromcusp}), which gives \cite{ZAMaldacena:1998im,ZARey:1998ik}:
\begin{equation}
 \alpha (\lambda )\simeq \frac{4\pi ^2\sqrt{\lambda }}{\Gamma ^4\left(\frac{1}{4}\right)}\qquad \left(\lambda \rightarrow \infty \right).
\end{equation}
The opposite limit $u_0\rightarrow \infty $, according to (\ref{ZAbremfromcusp}), yields the Bremsstrahlung function \cite{ZASemenoff:2004qr}:
\begin{equation}\label{ZABremsstrong}
 B(\lambda )\simeq \frac{\sqrt{\lambda }}{4\pi ^2}\qquad \left(\lambda \rightarrow \infty \right).
\end{equation}

The cusped Wilson loop or a generic wavy line cannot be computed by localization directly, because in general they do not preserve any supersymmetry. However, using  universality of the wavy line and the fact that some deformations of the circular Wilson loop are supersymmetric, one can use localization to compute the Bremstrahlung function  exactly  \cite{ZACorrea:2012at}.  Generalizations of these result to other observables and less supersymmetric theories have been studied in \cite{ZAFiol:2012sg,ZALewkowycz:2013laa,ZAFiol:2015spa,ZAFiol:2015mrp,ZAMitev:2015oty}.

The coupling to scalars that preserved enough supersymmetry for localization to apply is the latitude: $\mathbf{n}=(0,0,\cos\tau \,\sin\theta,\sin\tau \,\sin\theta ,\cos\theta ,0 )$, where $\theta $ is constant. The spacial part of the Wilson loop is the circle in the standard parameterization. The supersymmetry projectors (\ref{ZAbasicprojector}) for the latitude are of the form 
\begin{equation}
 \mathcal{P}^\pm=1\pm i\sin\theta \,\dot{x}^a\dot{x}^b\gamma _a\Gamma _b\gamma ^5\pm i\cos\theta \,\gamma ^0\gamma^1\Gamma _5x^a\gamma _a, 
\end{equation}
where in the last term we used the identities (\ref{ZAgamma-id}). Not all of the spinors $\bar{\epsilon }_0\mathcal{P}^-$ are of the form (\ref{ZAlinearspinor}) necessary for superconformal invariance, because of the middle term in the projector. We can get rid of this term by imposing an extra condition on $\bar{\epsilon }_0$:
\begin{equation}\label{ZAextra}
 \bar{\epsilon }_0\left(\gamma _3\Gamma _4+\gamma _4\Gamma _3\right)=0,
\end{equation}
or equivalently
\begin{equation}
 \bar{\epsilon }_0\gamma _{(a}\Gamma _{b)}=\frac{1}{2}\delta _{ab}\bar{\epsilon }_0
 \gamma ^c\Gamma _c.
\end{equation}
Then,
\begin{equation}\label{ZAlatitudeeps}
 \bar{\epsilon }\equiv \bar{\epsilon }_0\mathcal{P}^-
 =\bar{\epsilon }_0\left(1-\frac{i}{2}\,\sin\theta \,\gamma ^a\Gamma _a\gamma ^5
 -i\cos\theta \,\gamma ^0\gamma ^1\Gamma _5x^a\gamma _a\right),
\end{equation}
which is now a conformal Killing spinor.  The extra condition (\ref{ZAextra}) reduces the number of allowed supersymmetries by half, so the latitude is 1/4 BPS \cite{ZADrukker:2006ga}.

When $\theta =\pi /2$, the spinor (\ref{ZAlatitudeeps}) does not depend on $x^\mu $ at all, and the equatorial latitude is  invariant under 1/4 of the rigid supersymmetry \cite{ZAZarembo:2002an}. Its expectation value equals to one due to supersymmetry protection. For $\theta =0$ the contour on $S^5$ shrinks to a point,  and we get back to the circular Wilson loop discussed in sec.~\ref{ZAsec:CircularWL}.

As conjectured in \cite{ZADrukker:2006ga} and proved rigorously in \cite{ZAPestun:2009nn}, the exact expectation value of the latitude is given by the sum of rainbow diagrams for any $\theta $, not just $\theta =0$. The basic line-to-line propagator  is equal to
\begin{equation}
 \frac{\lambda }{8\pi ^2}\,\,\frac{|\dot{x}_1|\,|\dot{x}_2|\,
 \mathbf{n}_1\cdot \mathbf{n}_2
 -\dot{x}_1\cdot \dot{x}_2}{(x_1-x_2)^2}=\frac{\lambda \cos^2\theta }{16\pi ^2}\,,
\end{equation}
and is again a constant, rescaled by a factor of $\cos^2\theta $ compared to the circular loop case. The expectation value of the latitude is consequently given by the same expression (\ref{ZABessel}), under a simple replacement $\lambda \rightarrow \lambda \cos^2\theta $:
\begin{equation}\label{ZAexlat}
 \frac{1}{N}\,W(C_{\rm latitude})=\frac{2}{\sqrt{\lambda }\cos\theta }\,I_1\left(\sqrt{\lambda }\cos\theta \right).
\end{equation}

 At strong coupling:
\begin{equation}\label{ZAWlatude}
 W(C_{\rm latitude})\simeq \,{\rm e}\,^{\sqrt{\lambda }\cos\theta }.
\end{equation}
This result is in perfect agreement with the AdS/CFT duality. The minimal surface for the latitude \cite{ZADrukker:2005cu} is the direct product of the hemisphere (\ref{ZAcirclems}) in space-time and a solid angle with apex $2\theta $ on $S^5$ -- in the conformal gauge the solutions in $AdS_5$ and $S^5$ are independent provided each of them separately satisfies the Virasoro constraints. The regularized area of the hemisphere is $-2\pi $, while the solid angle subtended by the latitude is $+2\pi (1-\cos\theta )$, which altogether gives the area of $-2\pi \cos\theta $, to be multiplied by the string tension $\sqrt{\lambda }/2\pi $.  The exponent of the string amplitude that determines the expectation value of the latitude holographically  is thus exactly the same as the one in (\ref{ZAWlatude}).

The latitude with $\theta\rightarrow 0 $ can be regarded as a small perturbation of the circular Wilson loop. Even though the wavy line was originally defined by contour deformation in space-time, the quadratic part for the deviation on $S^5$  is controlled by the same Bremsstrahlung function  \cite{ZACorrea:2012at}:
\begin{equation}
 \frac{W(C_{\rm latitude})-W(C_{\rm circle})}{W(C_{{\rm circle}})}=-\frac{1}{2\pi ^2}\,B(\lambda )\theta ^2+\ldots 
\end{equation}
Since the expectation value for the latitude is obtained from that for the circle by replacing $\lambda \rightarrow \lambda \cos^2\theta\approx \lambda (1-\theta ^2) $, we get:
\begin{equation}
 B(\lambda )=\frac{1}{2\pi ^2}\,\,\frac{\partial \ln W(C_{\rm circle})}{\partial \ln\lambda }\,.
\end{equation}
The exact result for the circle (\ref{ZABessel}) then implies \cite{ZACorrea:2012at}:
\begin{equation}\label{ZAnon-pB}
 B(\lambda )=\frac{\sqrt{\lambda }I_2\left(\sqrt{\lambda }\right)}{4\pi ^2I_1\left(\sqrt{\lambda }\right)}\,.
\end{equation}
This is an exact result valid for arbitrary $\lambda $ and  large $N$ (the finite-$N$ result can be obtained by differentiating (\ref{ZALaguerre})). At strong coupling it agrees with the AdS/CFT prediction (\ref{ZABremsstrong}), since the ratio of the two Bessel functions approaches one when their arguments go to infinity.

Interestingly, the same function $B(\lambda )$ appears in the correlator of the straight Wilson line with the Lagrangian density operator inserted at infinity \cite{ZAFiol:2012sg}, thus reconfirming an interpretation of the Bremsstrahlung function in terms of the dipole radiation of an accelerated quark \cite{ZAMikhailov:2003er,ZACorrea:2012at,ZABanerjee:2015fed}. This relationship was recently elaborated further for $\mathcal{N}=2$ theories \cite{ZAFiol:2015spa,ZAFiol:2015mrp,ZAMitev:2015oty}.

Localization determines the leading order in the expansion of the cusp anomalous dimension around the supersymmetric configuration (the straight line). The other two limits considered above, $\phi \rightarrow \pi $ and $\phi \rightarrow i\infty $, are not supersymmetric. However, exact non-perturbative results are available even in this case, due to remarkable integrability properties  of the planar $\mathcal{N}=4$ SYM. In fact the whole function $\Gamma _{\rm cusp}(\phi ,\lambda )$ can be computed from Thermodynamic Bethe Ansatz equations  (TBA).
The light-like cusp is described by the asymptotic Bethe ansatz at any value of the coupling constant  \cite{ZABeisert:2006ez}, via its relationship to the twist-two local operators. The full machinery of TBA yields a set of more general functional equations which determine the cusp anomaly  at any $\phi $ and any $\lambda $ \cite{ZADrukker:2012de,ZACorrea:2012hh,ZAGromov:2015dfa,ZAGromov:2016rrp}.  The non-perturbative expression for the Bremsstrahlung function (\ref{ZAnon-pB}) can be recovered from the TBA equations, and can be generalized to include local operators inserted at the tip of the cusp \cite{ZAGromov:2012eu}.

 The latitude  belongs to a larger class of 1/8 supersymmetric Wilson loops, all of which can be computed by localization.
Suppose that the contour $C$ is restricted to lie on the surface of a two-dimensional sphere $S^2\subset \mathbb{R}^4$, such that at any $s$, $x^0(s)=0$, and $x^i(s)$ form a three-dimensional unit vector. The 1/8 BPS Wilson loop  
\cite{ZADrukker:2007dw} is then defined as
\begin{equation}\label{ZAS2WL}
 W_{2d}(C)=\left\langle \mathop{\mathrm{tr}}{\rm P}\exp
 \left[\oint_Cdx^i (iA_i+\varepsilon _{ijk}x^j\Phi ^k)\right]
 \right\rangle.
\end{equation}
It depends on three out of six scalar fields.
The supersymmetry projectors (\ref{ZAbasicprojector}) for this type of Wilson loops are
\begin{equation}
 \mathcal{P}^\pm=1\pm i\dot{x}^l\gamma _l\dot{x}^ix^j\varepsilon _{ijk}\gamma ^5\Gamma ^k.
\end{equation}
Using the identity
\begin{equation}
 \gamma _l=\frac{1}{2}\,\varepsilon _{lij}\gamma ^5\gamma ^0\gamma ^i\gamma ^j,
\end{equation}
the supersymmetry projector can be brought to a more concise form:
\begin{equation}
 \mathcal{P}^\pm = 1\pm i\gamma ^0\Gamma ^i\gamma _{ij}x^j.
\end{equation}

The transformation parameter $\bar{\epsilon }=\bar{\epsilon }_0\mathcal{P}^-$ is not really a Killing spinor (\ref{ZAlinearspinor}), unless extra conditions are imposed. The minimal set of conditions turns out to be 
\begin{equation}\label{ZAij}
 \bar{\epsilon }_0\left(\gamma _{ij}+\Gamma _{ij}\right)=0,
\end{equation}
 where indices $i$ and $j$ run from $1$ to $3$. Only two of these conditions are independent, because the gamma matrices used for the projection form a closed algebra under commutation. These conditions imply that
\begin{equation}
 \bar{\epsilon }_0\Gamma ^i=\frac{1}{3}\,\bar{\epsilon }_0\Gamma _j\gamma ^j\gamma ^i.
\end{equation}
The parameter of supersymmetry transformations  (\ref{ZAsuperspinor}) then becomes
\begin{equation}
 \bar{\epsilon }=\bar{\epsilon }_0\mathcal{P}^-=
 \bar{\epsilon }_0-\frac{2i}{3}\,\bar{\epsilon }_0\gamma ^0\Gamma _j\gamma ^jx^i\gamma _i,
\end{equation}
which is a superconformal Killing spinor. The two conditions (\ref{ZAij}) reduce the number of eligible constant spinors by a quarter, and the $\mathcal{P}^-$ projection by another half, so the Wilson loops defined in (\ref{ZAS2WL}) are indeed 1/8 BPS.

Quite remarkably, localization reduces the expectation values of the 1/8 BPS Wilson loops to Wilson loops in the bosonic two-dimensional Yang-Mills theory restricted to the zero-instanton sector \cite{ZADrukker:2007yx,ZADrukker:2007qr}. The path integral of $\mathcal{N}=4$ SYM localizes on two-dimensional field configurations
of the gauge field. The $S^2$ Wilson loops remain invariant under the action of the BRST operator used in localizing the path integral in this way \cite{ZAPestun:2009nn}.  The 4d and 2d coupling constants are related as $\lambda _{2d}=-\lambda /4\pi R^2$, where $R$ is the radius of the sphere. Since, the 2d coupling is negative, the localization partition function should be defined with care and requires complexification of the gauge group  \cite{ZAPestun:2009nn}.

\begin{figure}[t]
\begin{center}
 \centerline{\includegraphics[width=5cm]{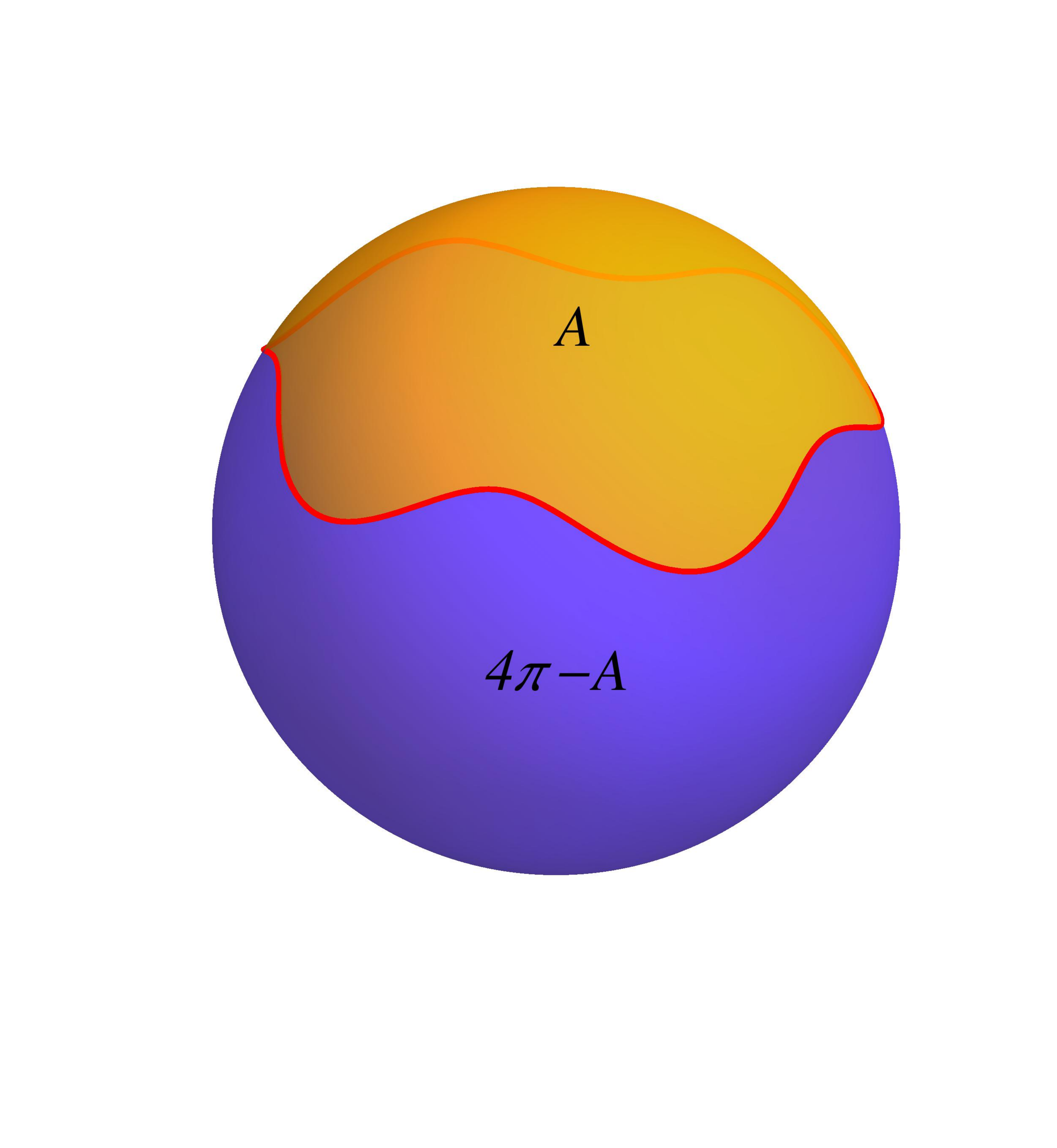}}
\caption{\label{ZAs2W}\small Wilson loop on $S^2$.}
\end{center}
\end{figure}

The two-dimensional Yang-Mills theory is invariant under area-preserving diffeomorphisms, so the Wilson loop without self-intersections can only depend on the area $A$  it encloses on $S^2$, in other words on  the solid angle at which the loop is seen from the middle of the sphere (fig.~\ref{ZAs2W}). It should also be symmetric under the interchange of the solid angle $A$ and its complement: $A\rightarrow 4\pi -A$. The exact expectation value of a general $S^2$ Wilson loop  is given by \cite{ZADrukker:2007yx,ZADrukker:2007qr}:
\begin{equation}
 \frac{1}{N}\,W_{2d}(C)=\frac{4\pi  }{\sqrt{\lambda A\left(4\pi -A\right)}}\,
 I_1\left(\frac{\sqrt{\lambda A\left(4\pi -A\right)}}{2\pi }\right).
\end{equation}
The latitude (\ref{ZAexlat}), for which  $A=2\pi (1-\sin\theta )$, is a particular example of this class of Wilson loops. 

\section{Operator Product Expansion}\label{ZAsec:OPE}

In addition to expectation values some correlation functions involving Wilson loops can also be computed with the help of localization. We will concentrate on the  two-point functions of Wilson loops with local gauge-invariant operators. In that case the problem can be reformulated in terms of the operator product expansion.

When probed from distances much larger than its size, a Wilson loop behaves as a local object, and can be approximated by a local operator insertion. This can be formalized by the operator product expansion of the loop operator \cite{ZAShifman:1980ui}:
\begin{equation}
 W(C,\mathbf{n})=\sum_{i}^{}\mathbb{C}_{i}[C,\mathbf{n}]\mathcal{O}_i(0),
\end{equation}
where $\mathcal{O}_i$ is a complete set of local gauge-invariant operators, and
$\mathbb{C}_{i}[C,\mathbf{n}]$ are numerical coefficients that depend on the shape of the contour $C$ and on the path $\mathbf{n}$ on the five-sphere. The OPE translates into an expansion of correlation functions of the Wilson loop in powers of $R/|x|$, where $R$ is the characteristic size of the  loop and $|x|$ is a typical scale of the problem. For instance,  a two-point function of a Wilson loop and a conformal primary scalar operator can be expanded as
\begin{equation}\label{ZAOPE}
 \left\langle W(C,\mathbf{n})\mathcal{O}_i(x)\right\rangle
 =\frac{\mathbb{C}_i[C,\mathbf{n}]}{|x|^{2\Delta _i}}+{\rm descendants},
\end{equation}
where $\Delta _i$ is the scaling dimension of $\mathcal{O} _i$. The contribution of descendants contains higher powers of $1/|x|$.

\begin{figure}[t]
\begin{center}
 \centerline{\includegraphics[width=4cm]{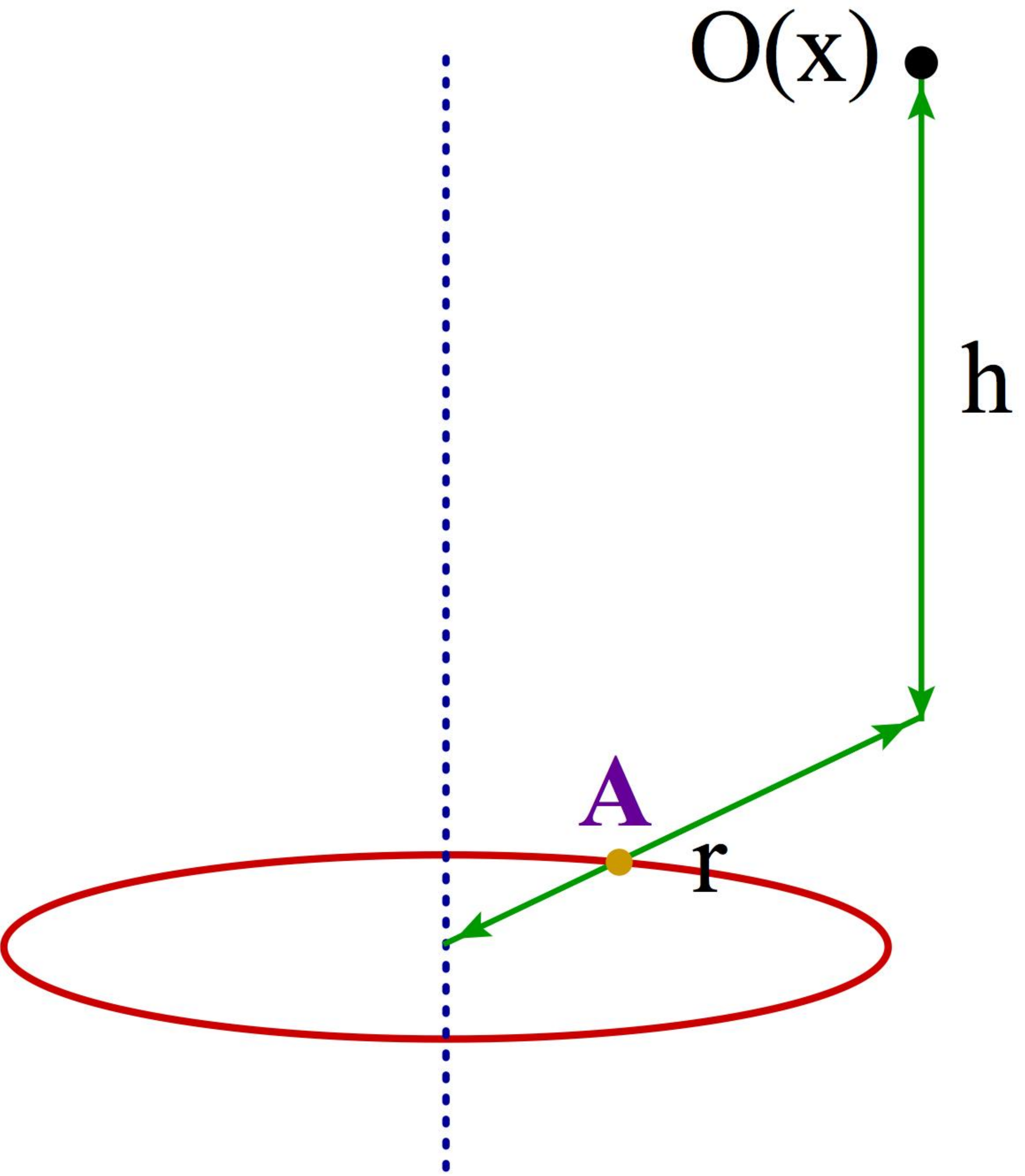}}
\caption{\label{ZAcirg}\small Correlator of the circular Wilson loop and a local operator.}
\end{center}
\end{figure}

Our basic example is the circular Wilson loop. In that case, the two-point correlator with a scalar primary is entirely determined by conformal symmetry, which is best seen after a conformal transformation that maps the circle to a line. In the setup illustrated in fig.~\ref{ZAcirg} this transformation is
an inversion centered at the point $A$. The correlator of a local operator and a Wilson line depends only on one length scale and therefore is fixed by scale invariance up to an overall constant. The inverse transformation then  determines the correlator with a circle. The overall constant can be identified with the OPE coefficient by matching to  (\ref{ZAOPE}) at large distances \cite{ZABerenstein:1998ij,ZAGomis:2008qa,ZAAlday:2011pf,ZABuchbinder:2012vr}: 
\begin{equation}
 \left\langle W(C_{\rm circle})\mathcal{O}_i(x)\right\rangle
 = \frac{\mathbb{C}_i}{\left[h^2+\left(r-R\right)^2\right]^{\frac{\Delta _i}{2}}\left[h^2+\left(r+R\right)^2\right]^{\frac{\Delta _i}{2}}}\,.
\end{equation}
Here $h$ is the distance from the point $x$ to the plane of the circle and $r$ the distance from $x$ to the circle's axis of symmetry (fig.~\ref{ZAcirg}). 

We are going to concentrate on the correlator of the circular loop with $\mathbf{n}=(1,\mathbf{0})$ and chiral primary operators (CPO):
\begin{equation}\label{ZACPOs}
 \mathcal{O}_J= \frac{1}{\sqrt{J}}\left(\frac{4\pi ^2}{\lambda }\right)^\frac{J}{2}
 \mathop{\mathrm{tr}}Z^J,
\end{equation}
where $Z=\Phi _1+i\Phi _2$. The chiral primaries are supersymmetry-protected and do not receive anomalous dimensions. The normalization factor is chosen such that the two-point function of $\mathcal{O}_J$ is unit-normalized:
\begin{equation}
 \left\langle \mathcal{O}^\dagger _J(x)\mathcal{O}_J(0)\right\rangle
 =\frac{1}{|x|^{2J}}\,.
\end{equation}

\begin{figure}[t]
\begin{center}
 \centerline{\includegraphics[width=4cm]{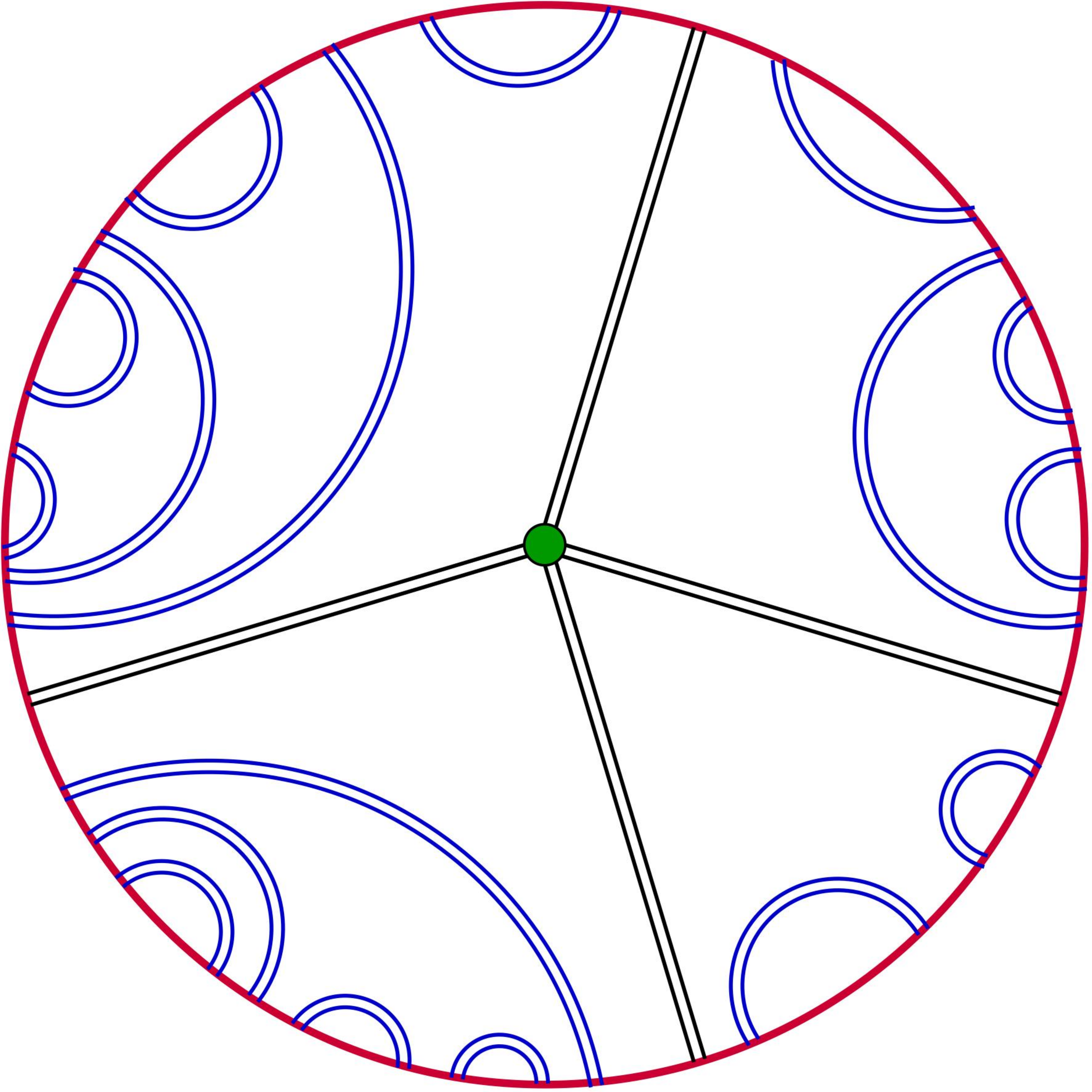}}
\caption{\label{ZArainbow-WO}\small The only diagrams that contribute to the correlation functions of the circular Wilson loop and a chiral primary operator are the rainbow diagrams without internal vertices.}
\end{center}
\end{figure}

The correlation functions $\left\langle W(C_{\rm circle})\mathcal{O}_J\right\rangle$ can be computed exactly using localization. The exact answer can again be obtained by summing the  rainbow diagrams  \cite{ZASemenoff:2001xp}, whereas a rigorous derivation relies on  localization of the path integral on $S^2$ \cite{ZAGiombi:2009ds}. 
The rainbow graphs now contain two types of propagators, those that connect the operator to the loop, and those that connect two different points on the loop, 
fig.~\ref{ZArainbow-WO}. These diagrams can be resummed by a brute-force account of combinatorics \cite{ZASemenoff:2001xp}. A more elegant derivation is based on mapping the problem to a Gaussian two-matrix model \cite{ZAGiombi:2009ms,ZABassetto:2009rt,ZAGiombi:2009ds,ZAGiombi:2012ep}.

Both types of propagators in fig.~\ref{ZArainbow-WO} are effectively constant: the loop-to-loop propagator is equal to $\lambda /16\pi ^2$, while the operator-to-loop propagator contributes a factor of $\lambda/8\pi ^2|x|^2$ for the operator  inserted far away from the loop. These two types of propagators are accounted for by  introducing two zero-dimensional fields, $\Phi $ and $Z$, with propagators

% \begin{equation}\label{ZAMatProps}
%   \wick[u]{1}{<1\Phi~~ >1\Phi  }=\frac{\lambda }{16^2}\,\qquad 
%   \wick[u]{1}{<1Z~~ >1\Phi  }=\frac{\lambda }{8\pi ^2i}
% \end{equation}

\begin{equation}\label{ZAMatProps}
\bcontraction{}{\Phi}{\,\,}{\Phi} \Phi \,\, \Phi =\frac{\lambda }{16^2}\,\qquad 
\bcontraction{}{Z}{\,\,}{\Phi} Z \,\, \Phi =\frac{\lambda }{8\pi ^2i}
\end{equation}

The factor of $i$ in the $Z\Phi $ propagator makes the quadratic form of the effective matrix model positive-definite. Since there are exactly $J$ $Z\Phi $ propagators in each diagram, this factor is easily  absorbed  into an overall normalization of the correlator. The necessity to introduce the factors of $i$ can be traced back to the fact that the 2d Yang-Mills theory, to which $\mathcal{N}=4$ SYM localizes, has a negative coupling and  requires complexification of the gauge group \cite{ZAPestun:2009nn}.

A Gaussian matrix integral that reproduces these propagators is
\begin{equation}\label{ZA2MMZ}
 \mathcal{Z}_{2MM}=\int_{}^{}dZ\,d\Phi\,\,{\rm e}\,^{-\frac{2\pi ^2N}{\lambda }\,\mathop{\mathrm{tr}}\left(Z^2+4iZ\Phi \right)}. 
\end{equation}
Integrating out $Z$ we get back to the matrix model (\ref{ZAN=4MM}) for the circular Wilson loop as expected.

The OPE coefficients  map to the following correlation function in the two-matrix matrix model (\ref{ZA2MMZ}):
\begin{equation}
 \mathbb{C}_J^{\rm CPO}=\frac{R^J}{\sqrt{J}}\left(-\frac{4\pi ^2}{\lambda }\right)^{\frac{J}{2}}\left\langle \mathop{\mathrm{tr}}Z^J\,\mathop{\mathrm{tr}}\,{\rm e}\,^{2\pi \Phi }\right\rangle.
\end{equation}
To calculate this correlator, we first get rid of one of the $Z$'s by Wick contracting it with a $\Phi $ in the exponential. The problem then reduces to computing  single-trace expectation values 
\begin{equation}\label{ZAMMmixed}
 W_k(s)=\left\langle \frac{1}{N}\,\mathop{\mathrm{tr}}\,{\rm e}\,^{s\Phi }Z^k\right\rangle,
\end{equation}
which are easier to deal with.
In terms of those,
\begin{equation}\label{ZApolustructure}
 \mathbb{C}_J^{\rm CPO}=\frac{\lambda R^J}{4\pi i\sqrt{J}}\left(-\frac{4\pi ^2}{\lambda }\right)^{\frac{J}{2}}W_{J-1}(2\pi ).
\end{equation}

To calculate the mixed correlator (\ref{ZAMMmixed})  we use the standard method of Schwinger-Dyson equations \cite{ZAMigdal:1984gj,ZAMakeenko:1991tb}. The Schwinger-Dyson equations for the two-matrix model (\ref{ZA2MMZ}) follow from the identity:
\begin{equation}
 \int_{}^{}dZ\,d\Phi \,\mathop{\mathrm{tr}}\left(\frac{\partial }{\partial \Phi ^t}\,
 \,{\rm e}\,^{s\Phi }Z^k\right)\,{\rm e}\,^{-\frac{2\pi ^2N}{\lambda }\,\mathop{\mathrm{tr}}\left(Z^2+4iZ\Phi \right)}=0,
\end{equation}
where $\partial /\partial \Phi ^t$ acts on everything to the right, making the integrand a total derivative. Performing differentiation we find:
\begin{equation}
\left\langle 
 \int_{0}^{s}dt\,\mathop{\mathrm{tr}}\,{\rm e}\,^{t\Phi }\,\mathop{\mathrm{tr}}\,{\rm e}\,^{\left(s-t\right)\Phi }Z^k-
 \frac{8\pi ^2iN}{\lambda }\,\mathop{\mathrm{tr}}\,{\rm e}\,^{s\Phi }Z^{k+1}
\right\rangle
=0.
\end{equation}
At large-$N$ the expectation value of the double trace appearing in the first term factorizes, and we  get a closed system of equations for the matrix-model loop amplitudes (\ref{ZAMMmixed}):
\begin{equation}\label{ZASch-D}
 W_{k+1}(s)=\frac{\lambda }{8\pi ^2i}\int_{0}^{s}dt\,W_0(t)W_k(s-t).
\end{equation}
A systematic way to solve these equations is to Laplace transform in $s$, which maps convolution to a product. We will not go through all the details, because the answer can be guessed after a number of easy sample computations.

The average without insertions $W_0(s)$ coincides with the expectation value of the circular loop (\ref{ZABessel}), up to a rescaling of the coupling constant:
\begin{equation}
 W_0(s)=\frac{4\pi }{s\sqrt{\lambda }}\,I_1\left(\frac{s\sqrt{\lambda }}{2\pi }\right),
\end{equation}
The Schwinger-Dyson equation   (\ref{ZASch-D}) can thus be viewed as a recursion relation that fixes $W_{k+1}$  
in terms of $W_k$.

The first step of recursion can be done with the help of the convolution formula for the Bessel functions: 
\begin{equation}\label{ZAconvB}
 \int_{0}^{a}\frac{dx}{x\left(a-x\right)}\,\,I_\mu \left(c(a-x)\right)I_\nu (cx)=
 \frac{\mu +\nu }{a\mu \nu }\,I_{\mu +\nu }\left(ca\right).
\end{equation}
For $W_1$ we then get the Bessel function again, but now with index two. The next iteration boils down to the same convolution formula, which produces  $I_3$, and so on. It is now easy to guess the general pattern:
\begin{equation}
 W_k(s)=\frac{2\left(k+1\right)}{is}\left(-\frac{\lambda }{4\pi ^2}\right)^{\frac{k-1}{2}}I_{k+1}\left(\frac{s\sqrt{\lambda }}{2\pi }\right).
\end{equation}
which can be straightforwardly checked to solve the recursion relations (\ref{ZASch-D}) by virtue of the convolution formula (\ref{ZAconvB}). 

Substituting the solution into (\ref{ZApolustructure}) we get a remarkably simple result for the OPE coefficient \cite{ZASemenoff:2001xp}:
\begin{equation}\label{ZAexactCPO}
 \mathbb{C}_J^{\rm CPO}=\sqrt{J}R^JI_J\left(\sqrt{\lambda }\right).
\end{equation}
This results holds at any $\lambda $, and at strong coupling can be compared to the predictions of the AdS/CFT correspondence. 

\begin{figure}[t]
\begin{center}
 \centerline{\includegraphics[width=13cm]{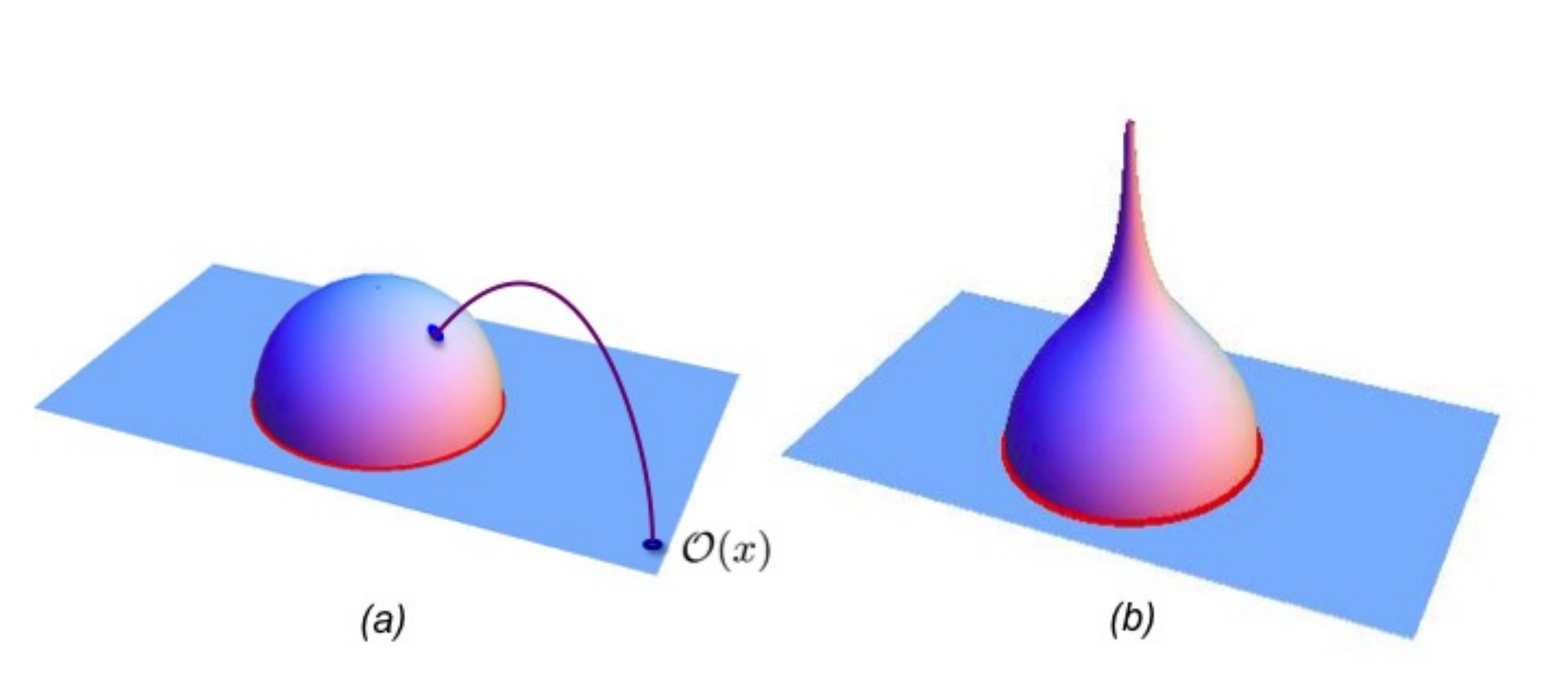}}
\caption{\label{ZAcstring}\small The correlation function $\left\langle W(C)\mathcal{O}(x)\right\rangle$ in string theory: (a) a bulk-to-boundary propagator stretched between the operator insertion and the string worldsheet, (b) emission of a  macroscopic string state (the local operator in this case is inserted at infinity).}
\end{center}
\end{figure}

In string theory, a local operator is dual to a closed string state and a Wilson loop  to a boundary state. In the most common situation the string dual of a local operator is well approximated by a supergravity field in the bulk. This is certainly true for the chiral primary operators (\ref{ZACPOs}) unless $J$ is parametrically large.
The correlation function $\left\langle W(C)\mathcal{O}(x)\right\rangle$ corresponds then to the following process: the operator insertion at the boundary emits a  supergravity mode  which is subsequently absorbed by the worldsheet created by the Wilson loop. This is illustrated in fig.~\ref{ZAcstring}a. When the operator is itself dual to a semiclassical string (an example is a CPO with $J\sim \sqrt{\lambda }$), the  whole process is described by a single worldsheet as shown in fig.~\ref{ZAcstring}b.

In general, the two-point function $\left\langle W(C,\mathbf{n})\mathcal{O}_i(x)\right\rangle$ (or the OPE coefficient $\mathbb{C}_i[C,\mathbf{n}]$, if the operator is placed at infinity) is computed by the string path integral (\ref{ZAWinstring}) with a vertex operator inserted. 
Dividing by the Wilson loop vev to normalize by the disc amplitude without insertions we get:
\begin{equation}\label{ZAdiscpf}
 \frac{\mathbb{C}_i[C,\mathbf{n}]}{W(C,\mathbf{n})}=\left\langle\int_{\Sigma }^{}d^2\sigma_o \,\sqrt{h}\, V_i(\sigma_o )\right\rangle,
\end{equation}
where the vertex operator $V_i(\sigma_o )$ represents the local operator $\mathcal{O}_i$ in SYM, and may depend on the string embedding coordinates $X^M(\sigma_o )$, their derivatives, worldsheet curvature, fermions and so on. 

The one-to-one map $V_i\longleftrightarrow \mathcal{O}_i$ is a core ingredient of the AdS/CFT duality, and yet it has never been worked out in any detail. Reason for that is a poor knowledge of string theory in $AdS_5\times S^5$. Not many vertex operators are actually known. The chiral primaries (\ref{ZACPOs}) constitute a fortunate exception. The string vertex operators, dual to CPOs,  can be calculated from the first principles, by expanding the string action in general supergravity fields around the $AdS_5\times S^5$ background \cite{ZABerenstein:1998ij}:
\begin{equation}\label{ZAcpovertex}
 V_J^{\rm CPO}=\frac{\left(J+1\right)\sqrt{J\lambda }}{8\pi N }
 \left(n_1+in_2\right)^Jz^J\left[
 \frac{\left(\partial x\right)^2-\left(\partial z\right)^2}{z^2}
 -\left(\partial \mathbf{n}\right)^2\right].
\end{equation}
 The dependence on $n_1+in_2$ and $z$ is dictated by the quantum numbers of the operator -- its R-charge and scaling dimension which are both equal to $J$. The normalization of the vertex operator and the structure of the second-derivative terms are dictated by the  AdS/CFT dictionary and by the couplings of the supergravity fields to the string worldsheet.

To calculate the OPE coefficient at the leading order in strong coupling it is enough to substitute the classical solution (\ref{ZAcirclems}) with constant $\mathbf{n}=(1,\mathbf{0})$ into the vertex operator (\ref{ZAcpovertex}) and integrate the latter over the worldsheet. The result  of this calculation  \cite{ZABerenstein:1998ij} is
\begin{equation}
 \frac{\mathbb{C}_J^{\rm CPO}}{W(C_{\rm circle})}\stackrel{\lambda \rightarrow \infty }{=}
 \frac{\left(J+1\right)R^J\sqrt{J\lambda }}{2N}\int_{0}^{\infty }d\tau \,\,\frac{\tanh^J\tau }{\cosh^2\tau }=\frac{R^J\sqrt{J\lambda }}{2N}\,.
\end{equation}
Taking into account that the ratio of the Bessel functions approaches one at infinity, we find that the string-theory calculation is in complete agreement with the exact results (\ref{ZAexactCPO}) and (\ref{ZABessel}).

Another tractable case is a BMN-like \cite{ZABerenstein:2002jq} limit in which $J$ goes to infinity simultaneously with $\lambda $ at fixed
\begin{equation}
 j=\frac{J}{\sqrt{\lambda }}\,.
\end{equation}
The backreaction of the vertex operator cannot be ignored in this case because of its exponential dependence on the large quantum number $J$. A heavy vertex operator  produces a source in the classical equations of motion of the sigma-model that distorts the shape of the macroscopic string worldsheet \cite{ZAPolyakovStrings2002}.

Since the vertex operator (\ref{ZAcpovertex}) carries an R-charge the string worldsheet will extend in $S^5$. In the parameterization $n_1+in_2=\cos\psi \,\,{\rm e}\,^{i\varphi }$, the string sitting at $\psi =0$ will maximize the weight in the path integral. The string action for the remaining degrees of freedom (in the conformal gauge) takes the form:
\begin{equation}\label{ZAstrwsources}
 S_{\rm str}=\frac{1}{2}\int_{}^{}d^2\sigma \,\left[
 \frac{\left(\partial x\right)^2+\left(\partial z\right)^2}{z^2}+\left(\partial \varphi  \right)^2
 \right]-2\pi j\ln z(\sigma _o)-2\pi ij\varphi (\sigma _o).
\end{equation}
 The equations of motion for $z$ and $\varphi $ acquire source terms, due to the vertex operator insertion:
\begin{eqnarray}
 &&-\partial ^2\varphi =2\pi ij\delta (\sigma -\sigma _o)
\nonumber \\
 && -\partial ^2\ln z-\frac{\left(\partial x\right)^2}{z^2}=2\pi j\delta \left(\sigma -\sigma _o\right)
\nonumber \\
&&-\partial ^a\left(\frac{\partial _ax^\mu }{z^2}\right)=0.
\nonumber 
\end{eqnarray}

The source terms produce singularities at $\sigma =\sigma _0$:
\begin{equation}
 \varphi \rightarrow -ij\ln|\sigma -\sigma _o|,
 \qquad 
 z\rightarrow \frac{\,{\rm const}\,}{|\sigma -\sigma _o|^j}
 \qquad \left(\sigma \rightarrow \sigma _o\right),
\end{equation}
which can be viewed as boundary conditions for the equations of motion.
The normalized OPE coefficient is given by the action $S_{\rm str}(j)$ evaluated on-shell:
\begin{equation}
 \frac{\mathbb{C}_J^{\rm CPO}}{W(C)}\simeq \,{\rm e}\,^{-\frac{\sqrt{\lambda }}{2\pi }\left(S_{\rm str}(j)-S_{\rm str}(0)\right)}.
\end{equation}
The $z$ and $\varphi $ parts of the string action (\ref{ZAstrwsources}) are separately log-divergent at the vertex operator insertion, but the total divergence actually cancels. This is a manifestation of marginality of the vertex operator.

It may seem that the solution is not unique, due to the dependence on the insertion point $\sigma _o$. But this is not the case. The insertion point is actually not arbitrary. The solution has to satisfy the Virasoro constraints, and this condition picks a unique $\sigma _o$. Alternatively one can start with (\ref{ZAdiscpf}), where $\sigma _o$ is an integration variable, and notice that at large $\lambda $ and $J$ the integral over $\sigma _o$ is semiclassical. Then $\sigma _o$ is determined by the saddle-point conditions. It can shown that the saddle-point equations on $\sigma _o$ are equivalent to the Virasoro constraints  \cite{ZAZarembo:2010rr}, again due to marginality of the vertex operator. 

The solution of the equations of motion for the circular Wilson loop was found in \cite{ZAZarembo:2002ph} and is most easily written in the exponential parameterization of the disc: $\sigma _0+i\sigma _1=\,{\rm e}\,^{-\tau +is}$. The vertex operator, for symmetry reasons, should be inserted at $\sigma =0$, or equivalently at $\tau =\infty $. Then,
\begin{eqnarray}
 \varphi &=&ij\tau
 \nonumber
  \\
 x^1+ix^2&=&\frac{\sqrt{j^2+1}\,{\rm e}\,^{j\tau +is}}{\cosh\left(
 \sqrt{j^2+1}\tau +\xi 
 \right)}
\nonumber \\
z&=&\,{\rm e}\,^{j\tau }\left[\sqrt{j^2+1}\tanh\left(\sqrt{j^2+1}\tau +\xi \right)-j\right]
\end{eqnarray}
with
\begin{equation}
 \xi =\ln\left(\sqrt{j^2+1}+j\right).
\end{equation}
The solution in shown in fig.~\ref{ZAcstring}b. The worldsheet has the shape of a funnel with an infinite spike that goes up to the horizon. The spike disappears once $j\rightarrow 0$  and the solution smoothly matches with the minimal surface (\ref{ZAcirclems}) for the circular Wilson loop. 

The action evaluated on the classical solution gives  \cite{ZAZarembo:2002ph}:
\begin{equation}
 \frac{\mathbb{C}_J^{\rm CPO}}{W(C_{\rm circle})}\simeq 
 \,{\rm e}\,^{-\sqrt{\lambda }\left[
 1-\sqrt{j^2+1}-j\ln\left(\sqrt{j^2+1}-j\right)
 \right]}.
\end{equation}
This is to be compared with the exact result (\ref{ZAexactCPO}) in which $J$ and $\lambda $ simultaneously go to infinity. The limit can be derived from the integral representation of the modified Bessel function:
\begin{equation}
 I_J\left(\sqrt{\lambda }\right)=\frac{\left(\frac{\sqrt{\lambda }}{2}\right)^J}{\sqrt{\pi }\Gamma \left(J+\frac{1}{2}\right)}
 \int_{-1}^{1}dt\,\left(1-t^2\right)^{J-\frac{1}{2}}\,{\rm e}\,^{\sqrt{\lambda }t}.
\end{equation}
For large $\lambda $ and $J$ the integral has a saddle point at $t=\sqrt{j^2+1}-j$, and with exponential accuracy:
\begin{equation}
 I_J\left(\sqrt{\lambda }\right)\simeq \,{\rm e}\,^{\sqrt{\lambda }
 \left[
 \sqrt{j^2+1}+j\ln\left(\sqrt{j^2+1}-j\right)
 \right]}.
\end{equation}
Normalization by the  expectation value of the Wilson (\ref{ZAStronglambdacircle}) brings this result into the full agreement with the string-theory calculation.

Localization allows one to study much wider class of correlation functions involving  Wilson loops of different shape \cite{ZAGiombi:2009ds,ZAEnari:2012pq}, in higher representations of the gauge group \cite{ZAGomis:2008qa},  't~Hooft loops \cite{ZAGomis:2008qa,ZAGiombi:2009ek,ZABassetto:2010yc}, correlators of two Wilson loops \cite{ZAGiombi:2009ms,ZABassetto:2009rt,ZABassetto:2009ms,ZAGiombi:2012ep} as well as multi-point correlation functions \cite{ZAGiombi:2012ep}.

\section{Massive theory}\label{ZAsec:massive}

A minimal amount of supersymmetry sufficient to localize a path integral on $S^4$ is $\mathcal{N}=2$ \cite{ZAPestun:2007rz}. While $\mathcal{N}=4$ SYM is unique, there are many $\mathcal{N}=2$ gauge theories and their localization partition functions  have qualitatively new features compared to the $\mathcal{N}=4$ case. The resulting matrix models are not Gaussian any more, and there is no simple map between Feynman diagrams and the matrix integral\footnote{It is interesting, in this respect, to compare explicit perturbative calculations  in the $\mathcal{N}=2$ superconformal QCD \cite{ZAAndree:2010na} with localization. A three-loop propagator correction, the first diagram that goes beyond the rainbow approximation \cite{ZAAndree:2010na}, can be identified in the matrix model \cite{ZAPasserini:2011fe}. This correction involves a $\zeta (3)$ transcendentality  from the loop integration, while in the matrix model $\zeta (3)$ appears directly in the action (see \cite{ZAMitev:2015oty} for further discussion of transcendental numbers appearing in the localization formulas and their comparison to perturbation theory).}. The instantons, that did not contribute to the partition function of $\mathcal{N}=4$ SYM, survive localization in generic $\mathcal{N}=2$ theories. At large-$N$ the instantons are exponentially suppressed and will actually be neglected in what follows. Finally, and perhaps most interestingly, localization does not rely on conformal symmetry and applies to massive theories as well.

Breaking supersymmetry and introducing a mass scale in holographic duality is  conceptually simple. A feature in the bulk (typically a domain wall or a black hole horizon) distance $z_0=1/M$ away from the boundary sets the mass scale $M$ in the dual gauge theory. Difficulties lie in formulating the holographic dictionary at the string level, which requires the resulting geometry to be a consistent string background. Perhaps the most reliable approach is to start with $\mathcal{N}=4$ SYM deformed by  a relevant operator. The string dual then is a continuous deformation of $AdS_5\times S^5$. Switching on a relevant perturbation corresponds to imposing boundary conditions and evolving the bulk fields according to the supergravity equations of motion away from the boundary. The only relevant deformation of $\mathcal{N}=4$  SYM that preserves $\mathcal{N}=2$ supersymmetry  is known as the $\mathcal{N}=2^*$ theory. The dual supergravity background is known explicitly in this case \cite{ZAPilch:2000ue}. 

The  $\mathcal{N}=2$ decomposition of the $\mathcal{N}=4$ supermultiplet consists of the  vector multiplet, containing the gauge fields $A_\mu $, two scalars $\Phi $ and $\Phi '$ and two Majorana fermions, and two CPT conjugate hypermultiplets, containing two complex scalars $Z_i$ and a Dirac fermion. The only relevant operator that
 one can add to the original $\mathcal{N}=4$ Lagrangian without breaking $\mathcal{N}=2$ supersymmetry is the mass term for the hypermultiplet\footnote{In components, it yields the dimension-2 $\bar{Z}^iZ_i$ mass term, the dimension-3 mass term for the Dirac fermion, the $\Phi '\varepsilon ^{ij}\mathop{\mathrm{Im}}(Z_iZ_j)$ trilinear coupling, which breaks symmetry between $\Phi $ and $\Phi '$, and certain Yukawa couplings.}. 
  
 The path integral of the $\mathcal{N}=2^*$ theory compactified  on $S^4$ localizes to the following eigenvalue model \cite{ZAPestun:2007rz}:
\begin{equation}\label{ZAN=2star}
 Z=\int_{}^{}d^N a\,\,\prod_{i<j}^{}\frac{\left(a_i-a_j\right)^2H^2\left(a_i-a_j\right)}{H\left(a_i-a_j+M\right)H\left(a_i-a_j-M\right)}\,\,{\rm e}\,^{-
 \frac{8\pi ^2N}{\lambda }\,\sum\limits_{i}^{}a_i^2
 },
\end{equation}
where the function $H(x)$ is defined by an infinite product
\begin{equation}
 H(x)=\prod_{n=1}^{\infty }\left(1+\frac{x^2}{n^2}\right)^n\,{\rm e}\,^{-\frac{x^2}{n}}.
\end{equation}
We have neglected instantons, keeping in mind that they are suppressed in the large-$N$ limit.
The integration variables are the eigenvalues of the zero mode of the scalar $\Phi $:
\begin{equation}\label{ZAHiggs}
 \Phi =\mathop{\mathrm{diag}}\left(a_1,\ldots ,a_N\right).
\end{equation}

The expectation value of the Wilson loop along the big circle of $S^4$ is given by the same formula (\ref{ZAMMWL}), provided the original Wilson loop operator couples exactly to the same scalar. Because the theory at hand is not conformal any more, the circular loop on $S^4$ cannot be mapped back to $\mathbb{R}^4$. The dependence on the radius $R$ of $S^4$ also does not scale away. For brevity we have set $R=1$, so dimensionful quantities such as $M$, $\Phi $ and $a_i$ should be  understood as $MR$, $\Phi R$ and $a_iR$.

\begin{figure}[t]
\begin{center}
 \centerline{\includegraphics[width=10cm]{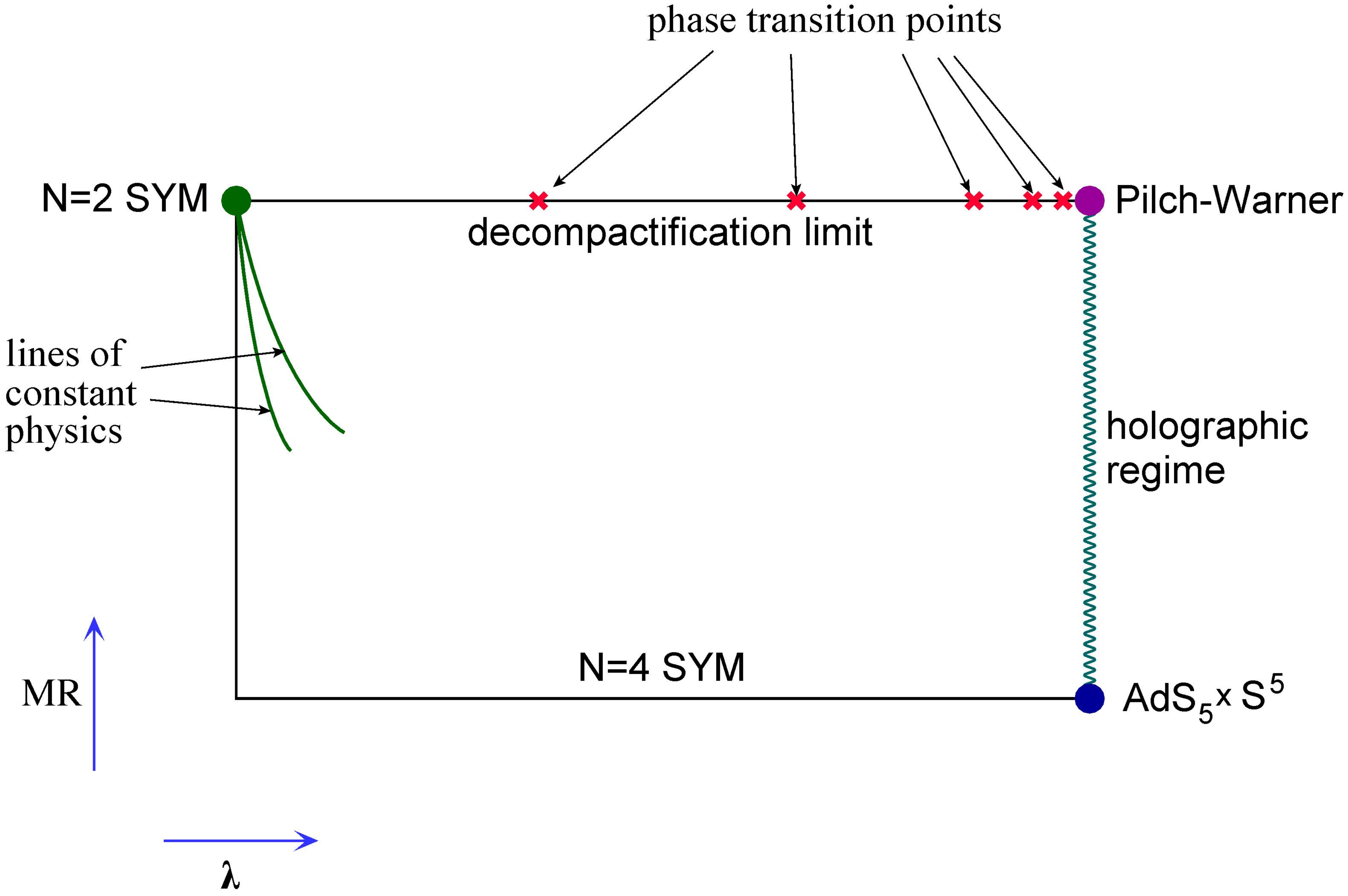}}
\caption{\label{ZAPhD}\small The phase diagram of $\mathcal{N}=2^*$ theory on $S^4$ (from \cite{ZARusso:2013kea}).}
\end{center}
\end{figure}

The saddle-point equations for the eigenvalue model (\ref{ZAN=2star}) are
\begin{equation}\label{ZAsaddleNstar}
 \frac{1}{N}\sum_{j\neq i}^{}
 \left(
 \frac{1}{a_i-a_j}-\mathcal{K}(a_i-a_j)+\frac{1}{2}\,\mathcal{K}(a_i-a_j+M)
 +\frac{1}{2}\,\mathcal{K}(a_i-a_j-M)
 \right)
 =\frac{8\pi ^2}{\lambda }\,a_i,
\end{equation}
where 
\begin{equation}
 \mathcal{K}(x)=-\frac{H'(x)}{H(x)}\,.
\end{equation}
These equations were studied in \cite{ZARusso:2012kj,ZARusso:2012ay,ZABuchel:2013id,ZARusso:2013qaa,ZARusso:2013kea,ZAChen:2014vka,ZAZarembo:2014ooa}, and although their general solution is not known, the phase diagram in the $(M,\lambda )$ plane can be mapped in a fair amount of detail, and turns out to be rather non-trivial, fig.~\ref{ZAPhD}. 

When $M\rightarrow \infty $ and $\lambda \rightarrow 0$ simultaneously, the hypermultiplets can be integrated out leaving behind pure $\mathcal{N}=2$ SYM. The mass scale  $M$ plays the r\^ole of a UV cutoff in the low-energy theory, while $\lambda $ is identified with the bare coupling. The beta-function of the $\mathcal{N}=2$ SYM then generates a dynamical scale $\Lambda =M\,{\rm e}\,^{-4\pi ^2/\lambda }$ (green lines in fig.~\ref{ZAPhD} are  the lines of constant $\Lambda $). The saddle-point equations of the localization matrix model reproduce \cite{ZARusso:2012ay} in this corner of the phase diagram the large-$N$ solution of $\mathcal{N}=2$ SYM, known  from the Seiberg-Witten theory \cite{ZADouglas:1995nw,ZAFerrari:2001mg}.

\begin{figure}[t]
\begin{center}
 \centerline{\includegraphics[width=8cm]{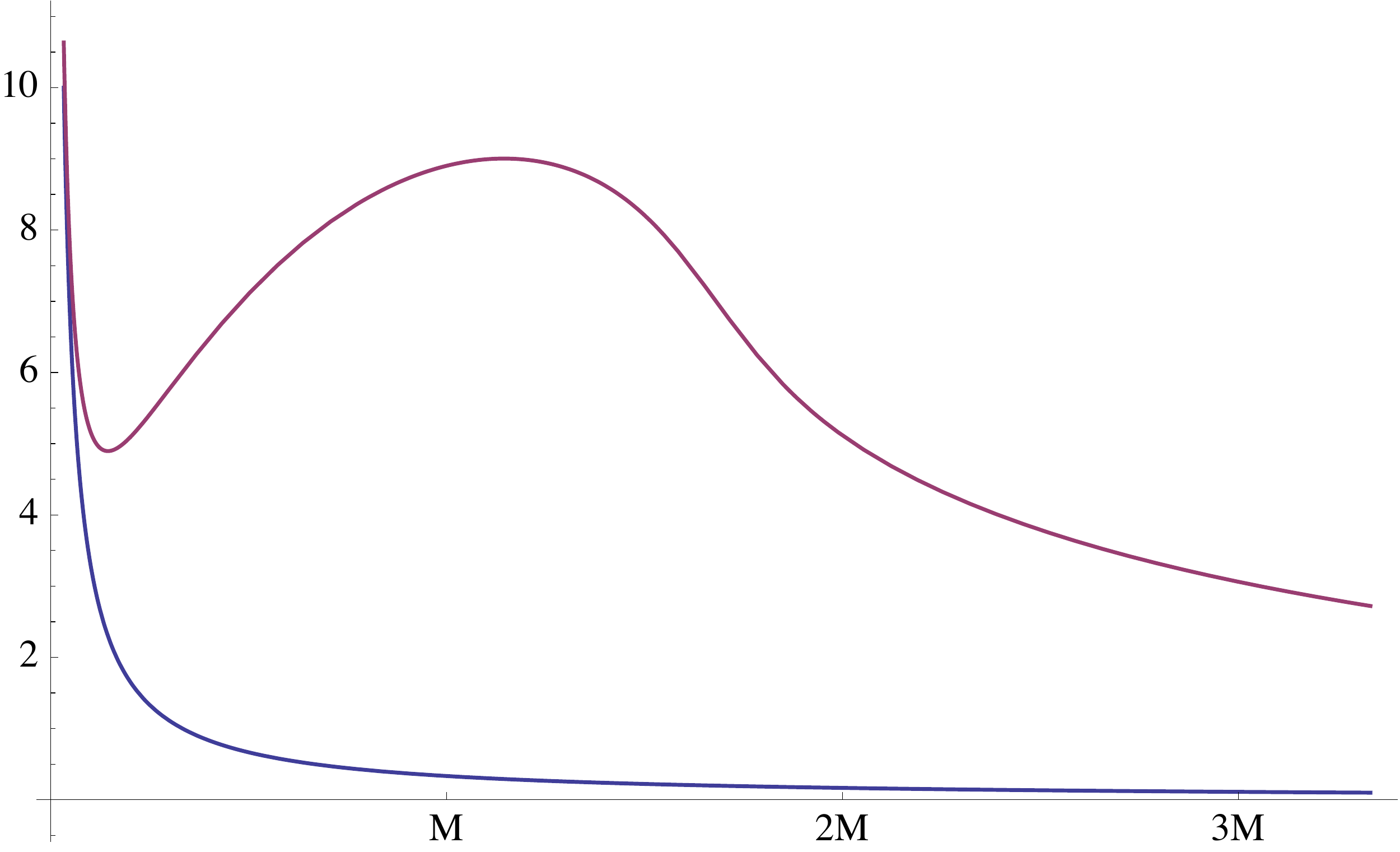}}
\caption{\label{ZAeforce}\small The two-body force acting between the eigenvalues in the $\mathcal{N}=2^*$ localization matrix model for $M=3$ (upper curve) and in the Gaussian model (lower curve).}
\end{center}
\end{figure}

The one-body potential in the $\mathcal{N}=2^*$ matrix  model  is still Gaussian, while the two-body potential gets modified by the mass deformation. The two-body force between eigenvalues has a rather intricate shape (fig.~\ref{ZAeforce}). Remaining universally repulsive, it does not decrease with distance as fast as in the Gaussian model, and can compete with the attractive one-body potential. This competition  causes  an infinite sequence of quantum phase transitions in the decompactification limit $R\rightarrow \infty $ (which in the dimensionless variables that we use corresponds to $M\rightarrow \infty $) \cite{ZARusso:2013qaa}\footnote{The phase transitions happen at $\lambda _c^{(1)}=35.42...$, $\lambda _c^{(2)}=84.6\pm 1.0$, $\lambda _c^{(3)}=153.0\pm 0.7$, and asymptotically at $\lambda _c^{(n)}\simeq \pi ^2n^2$. The first critical coupling is known exactly \cite{ZARusso:2013qaa}. The numerical results for secondary transitions improve on estimates of \cite{ZARusso:2013qaa} and are obtained with the help of the formalism developed in \cite{ZAZarembo:2014ooa}.}. Physically the phase transitions arise because of the resonances on nearly massless hypermultiplets. Indeed, the masses of the hypermultiplet fields in the Higgs background (\ref{ZAHiggs}) are $m^{\rm h}_{ij}=|a_i-a_j\pm M|$ and can become small if the distance between a pair of eigenvalues gets close to $M$.

Since we mainly focus on an interplay between localization and holography, we are interested in  the strong-coupling limit of $\mathcal{N}=2^*$ SYM \cite{ZABuchel:2013id,ZAChen:2014vka,ZAZarembo:2014ooa}. Drawing intuition from the solution of the Gaussian model  (\ref{ZAWigner}), (\ref{ZAmuGauss}), we may assume that the width of the eigenvalue distribution grows with $\lambda $ and will be much larger than any other scale in the problem at strong coupling. This is certainly true for small $M$, and will be checked {\it a posteriori} for arbitrary $M$. Treating $M $ as a small parameter, and using the large-distance asymptotics of the function $\mathcal{K}(x)$, we find:
$$
\frac{1}{2}\,\mathcal{K}(x+M)+\frac{1}{2}\,\mathcal{K}(x-M)-\mathcal{K}(x)\approx M^2\mathcal{K}''(x)\approx \frac{M^2}{x}
$$
Hence only the tail of the two-body force in fig.~\ref{ZAeforce} is important at strong coupling, and its sole effect is to renormalize the $1/x$ interaction of the Gaussian model. The saddle-point distribution then obeys the Wigner law (\ref{ZAWigner}) with \cite{ZABuchel:2013id}
\begin{equation}\label{ZALOstrongmu}
 \mu =\frac{\sqrt{\lambda \left(M^2+\frac{1}{R^2}\right)}}{2\pi }\,.
\end{equation}
We have re-instated the dependence on $R$ and the canonical mass dimension of $\mu $ and $M$. When $M$=0, there are no dimensionful parameters in the problem and $\mu $ scales away as $1/R$, while in the $\mathcal{N}=2^*$ theory it freezes at the scale that is parametrically larger than the bare mass in the Lagrangian, in accord with our original assumption.

The strong-coupling asymptotics of the circular Wilson loop is governed  by the largest eigenvalue: $W(C_{\rm circle})\simeq \,{\rm e}\,^{\sqrt{\lambda }MR}$. 
Although we cannot calculate any Wilson loop apart from the circle, it is natural to assume that expectation values for sufficiently large loops are universal, and hence should obey  perimeter law with the coefficient fixed by localization:
\begin{equation}\label{ZAperlaw}
 W(C)\simeq \,{\rm e}\,^{\frac{\sqrt{\lambda }}{2\pi }\,L(C)}.
\end{equation}

This prediction can be checked using the explicit form of the dual supergravity background \cite{ZABuchel:2013id}. The relevant part of the metric is \cite{ZAPilch:2000ue}
\begin{equation}
 ds^2=\frac{AM^2}{c^2-1}\,dx_\mu ^2+\frac{1}{A\left(c^2-1\right)^2}\,dc^2,
 \qquad 
 A=c+\frac{c^2-1}{2}\,\ln\frac{c-1}{c+1}\,.
\end{equation}
The holographic coordinate $c$ is related to $z$ in (\ref{ZAPoincare}) by
\begin{equation}
 c=1+\frac{z^2M^2}{2}+\ldots 
\end{equation}
so the boundary is at $c=1$. One can check that near the boundary the metric  indeed asymptotes to that of $AdS_5$. 

The minimal surface for a sufficiently big contour is approximately a cylinder, repeating the shape of the Wilson loop for any $c$, as long as $c\ll ML$. Because the metric decreases with $c$ very fast, most part of the area will come from this region, and we can neglect the bending and eventual closure of the minimal surface in computing the area:
\begin{equation}
 A_{\rm min}(C)=ML\int_{1+\frac{M^2\varepsilon ^2}{2}}^{\infty }
 \frac{dc}{\left(c^2-1\right)^{\frac{3}{2}}}=\frac{L}{\varepsilon }-ML.
\end{equation}
The divergent term is subtracted by regularization and, taking into account that the dimensionless string tension must be the same as in $AdS_5\times S^5$, $T=\sqrt{\lambda }/2\pi $, we get perimeter law with exactly the same coefficient  (\ref{ZAperlaw}) as inferred from localization.

As shown in \cite{ZABobev:2013cja} the free energy of the matrix model agrees with the on-shell action of the supergravity on the solution that has $S^4$ as a boundary. Corrections in $1/\sqrt{\lambda }$ to the leading-order strong-coupling result (\ref{ZALOstrongmu}) have been calculated on the matrix model  side \cite{ZAChen:2014vka,ZAZarembo:2014ooa}, and it would be very interesting to compare them to  quantum corrections due to string fluctuations in the bulk. 

\section{Conclusions}\label{ZAsec:Conclusions}

Localization is a powerful tool to explore supersymmetric gauge theories in the non-perturbative domain. Although limited to a restricted class of observables, localization relies on a direct evaluation of the path integral, without recourse to any assumptions or uncontrollable approximations. Via holography these first-principle calculations can be confronted with string theory and can give us additional hints on how string description emerges form summing planar diagrams.

Localization predictions are sometimes rather detailed. This review focusses on just a few examples, and in particular leaves aside theories for which a holographic dual is not really well established or has no weakly coupled regime. One interesting example of this class is $\mathcal{N}=2$ superconformal QCD -- an $\mathcal{N}=2$ supersymmetric gauge theory with $N_f=2N_c$ fundamental hypermultiplets. This theory has zero beta-function, and is presumably dual to strings on $AdS_5\times X_5$, where $X_5$ may not be geometric (see \cite{ZAGadde:2009dj} for a concrete proposal). The strong-coupling solution of the matrix model for $\mathcal{N}=2$ super-QCD is very different from the  $\mathcal{N}=4$ and $\mathcal{N}=2^*$ cases  \cite{ZAPasserini:2011fe}. Potential implications of this result for holography have not been worked out so far. Another class of examples are two-dimensional theories with $\mathcal{N}=4$ supersymmetry, which are dual to strings on $AdS_3\times S^3\times T^4$ supported by the RR flux. Here on the contrary the planar diagram expansion on the gauge-theory side is not easy to develop (see \cite{ZAPakman:2009zz} and \cite{ZASax:2014mea} for two different proposals). Localization on $S^2$ \cite{ZABenini:2012ui,ZADoroud:2012xw} (see \volcite{BL}) may be very useful in this respect, and it would be interesting to solve the resulting matrix model at large-$N$.

An interplay between holography and localization has been studied in much detail in three dimensions (see \cite{ZAMarino:2011nm} for a review), and in dimensions higher than four \cite{ZAKallen:2012zn,ZAJafferis:2012iv,ZAMinahan:2013jwa,ZAMinahan:2015jta,ZAMinahan:2015any}. It is also possible to localize on manifolds different from $S^4$ (see \volcite{HO}), which has a number of interesting applications to holography. Entanglement entropy of a spherical region can be computed that way \cite{ZAHuang:2014pda,ZACrossley:2014oea} and compared to the Ryu-Takayanagi prescription \cite{ZARyu:2006bv}  at strong coupling.  Bremsstahlung function in generic $\mathcal{N}=2$ theories can be also extracted from localization \cite{ZAFiol:2015spa}. Localization of $\mathcal{N}=4$ SYM on a large class of manifolds of the form $S^1\times M_3$ \cite{ZAAssel:2014paa} yields supersymmetric indices that can be compared \cite{ZAGenolini:2016sxe} to the supergravity action on geometries found in  \cite{ZACassani:2014zwa}.

\subsection*{Acknowledgements}
I would like to thank X.~Chen-Lin, A.~Dekel, N.~Drukker, J.~Gordon, D.~Medina Rincon, J.~Russo and G.~Semenoff for many useful discussions on localization, AdS/CFT correspondence, and matrix models.
This work was supported by the Marie
Curie network GATIS of the European Union's FP7 Programme under REA Grant
Agreement No 317089, by the ERC advanced grant No 341222, by the Swedish Research Council (VR) grant
2013-4329, and by RFBR grant 15-01-99504.

\documentfinishBBL
% \ifx\ifLONG\undefined
% % if using bibtex uncomment the next line and put ZA.bib file
% %\bibliographystyle{bibreview} \bibliography{ZA} 
% \input{ZA.long.bbl}
% \end{document} 
% \else
% \input{Zarembo/ZA.short.bbl}
% \fi 